\documentclass[12pt]{article}
\usepackage{amssymb,amsmath,mathtools,mathrsfs,slashed}
\usepackage{epsfig}
\usepackage{color,graphicx}
\usepackage{cite}
\usepackage{amssymb,amsmath}
\usepackage{multirow}
\usepackage{array}
\usepackage{hyperref}
\usepackage{comment}
\usepackage{epstopdf}

\newcommand{\rc}{\nonumber\\}

\newcommand{\be}{\begin{equation}}
\newcommand{\ee}{\end{equation}}
\newcommand{\bea}{\begin{eqnarray}}
\newcommand{\eea}{\end{eqnarray}}

\newcommand{\ruv}{\ensuremath{R_{_{UV}}}}
\newcommand{\cuv}{\ensuremath{c_{_{UV}}}}

\usepackage[top=2cm, bottom=2cm, left=2cm, right=2cm]{geometry}	
\numberwithin{equation}{section}

\begin{document}
 
\begin{flushright}
HIP-2019-39/TH
\end{flushright}

\begin{center}

\centering{\Large {\bf Holographic spontaneous anisotropy}}

\vspace{8mm}

\renewcommand\thefootnote{\mbox{$\fnsymbol{footnote}$}}
Carlos Hoyos,${}^{1,2}$\footnote{hoyoscarlos@uniovi.es},
Niko Jokela${}^{3,4}$\footnote{niko.jokela@helsinki.fi},\\
Jos\'e Manuel Pen\'\i n${}^{5,6,7}$\footnote{jmanpen@gmail.com}, and
Alfonso V. Ramallo${}^{5,6}$\footnote{alfonso@fpaxp1.usc.es},

\vspace{8mm}
${}^1${\small \sl Department of Physics}  and\\
${}^2$ {\small \sl Instituto de Ciencias y Tecnolog\'{\i}as Espaciales de Asturias (ICTEA)} \\
{\small \sl Universidad de Oviedo} \\
{\small \sl E-33007/33004, Oviedo, Spain}

\vspace{4mm}
${}^3${\small \sl Department of Physics} and ${}^4${\small \sl Helsinki Institute of Physics} \\
{\small \sl P.O.Box 64} \\
{\small \sl FIN-00014 University of Helsinki, Finland}

\vspace{2mm}
\vskip 0.2cm

${}^5${\small \sl Departamento de  F\'\i sica de Part\'\i  culas} 
{\small \sl and} \\
${}^6${\small \sl Instituto Galego de F\'\i sica de Altas Enerx\'\i as (IGFAE)} \\
{\small \sl Universidade de Santiago de Compostela} \\
{\small \sl E-15782 Santiago de Compostela, Spain} 

\vspace{2mm}
\vskip 0.2cm

${}^7${\small \sl Mathematical Sciences and STAG Research Center,} \\ 
{\small \sl University of Southampton, Highfield, Southampton SO17 1BJ, UK}

\vskip 0.2cm
\vskip 0.2cm

\end{center}

\vspace{8mm}

\setcounter{footnote}{0}
\renewcommand\thefootnote{\mbox{\arabic{footnote}}}

\begin{abstract}
\noindent 
We construct a family of holographic duals to  anisotropic states in a strongly coupled gauge theory. On the field theory side the anisotropy is generated by giving a vacuum expectation value to a dimension three operator. We obtain our gravity duals by considering the geometry corresponding to the intersection of D3- and D5- branes along 2+1 dimensions. Our backgrounds are supersymmetric and solve the fully backreacted equations of motion of ten-dimensional supergravity with smeared D5-brane sources. In all cases the geometry flows to $AdS_{5}\times {\mathbb S}^5$ in the UV, signaling an isotropic UV fixed point of the dual field theory. In the IR, depending on the parameters of the solution, we find two possible behaviors: an isotropic fixed point or a geometry with anisotropic Lifshitz-like scaling symmetry.  We study several properties of the solutions, including the entanglement entropy of strips.  We show that any natural extension of existing $c$-functions will display non-monotonic behavior, conforming with the presence of new degrees of freedom only at intermediate energy scales.
\end{abstract}

\newpage
\tableofcontents


\newpage


\section{Introduction}

A holographic description of anisotropic but homogeneous phases of strongly coupled theories is interesting for its potential application to a varied set of systems in high energy physics and condensed matter physics. 

In the context QCD, the initial stages of the quark-gluon plasma formed in heavy ion collisions are highly anisotropic due to the initial conditions. The effect of the initial anisotropy on the properties of the quark-gluon plasma using holography was first studied in  \cite{Janik:2008tc,Chesler:2008hg}.  Anisotropic phases could also appear in cold but dense matter such as the one found in the interior of neutron stars, especially in the presence of strong magnetic fields. This in principle could lead to the observation of stars more compact than the ones allowed by isotropic matter, see, {\emph{e.g.}}, \cite{Yagi:2015upa,Yagi:2016ejg}. Strongly coupled anisotropic phases have been studied using holography in a variety of setups, including axionic/dilatonic sources \cite{Mateos:2011ix,Mateos:2011tv,Koga:2014hwa,Jain:2014vka,Banks:2015aca,Roychowdhury:2015cva,Roychowdhury:2015fxf,Misobuchi:2015ioa,Banks:2016fab,Avila:2016mno,Donos:2016zpf,Giataganas:2017koz,Itsios:2018hff,Arefeva:2018hyo,Liu:2019npm,Inkof:2019gmh}, electric \cite{Karch:2007pd,Albash:2007bk,Albash:2007bq} and magnetic fields \cite{Erdmenger:2007bn,DHoker:2009mmn,DHoker:2009ixq,Jensen:2010vd,Evans:2010hi,DHoker:2010zpp,Kim:2010pu,Hoyos:2011us,Ammon:2012qs,Gursoy:2018ydr}  or both \cite{Evans:2011tk,Kharzeev:2011rw,Donos:2011pn,Jokela:2015aha,Itsios:2016ffv},  and $p$-wave superfluids \cite{Gubser:2008wv,Ammon:2008fc,Basu:2008bh,Iizuka:2012iv,Donos:2012gg,Iizuka:2012wt}. Strongly coupled holographic matter has also been studied in the context of compact stars \cite{Hoyos:2016zke,Annala:2017tqz,Jokela:2018ers,Ishii:2019gta,Chesler:2019osn,Hirayama:2019vod,Ecker:2019xrw,Fadafa:2019euu}, so a combination of the approaches will lead us to a fascinating unknown territory, the ``mass-gap'' between the heaviest neutron stars and the lightest black holes.

In systems with strongly correlated electrons, anisotropic nematic phases appear in the presence of magnetic fields in ultra-clean quantum Hall systems and in Sr${}_3$Ru${}_2$O${}_7$, and there is evidence that similar phases are present in iron-based and cuprate high $T_c$ superconductors (see \cite{Fradkin:2010rev} for a review on the topic). The application of holography to anisotropic and multilayered condensed matter systems has produced many interesting results \cite{Penin:2017lqt,Jokela:2019tsb,Gran:2019djz}.

Among one of the most surprising observations in holographic duals with broken spatial symmetries is the existence of `boomerang' flows \cite{Donos:2017ljs,Donos:2017sba}, where the renormalization group (RG) flow drives the theory in the far UV and far IR to isotropic fixed points with the same number of degrees of freedom, as counted by the holographic $c$-function \cite{Freedman:1999gp}. This seems at odds with the usual intuition of Wilsonian flow where the number of degrees of freedom is reduced by coarse graining as one moves from higher to lower energy scales. Nevertheless, the non-monotonicity is not in contradiction with any of the existing $c$-theorems \cite{Zamolodchikov:1986gt,Freedman:1999gp,Casini:2004bw,Myers:2010xs,Myers:2010tj,Komargodski:2011vj,Casini:2012ei}, as all rely on Lorentz invariance to prove the existence of a monotonic quantity under the RG flow evolution. In principle, a similar measuring device  may not exist in an anisotropic flow (or be a very complicated object) even if the Wilsonian intuition is correct. An interesting question is whether boomerang flows are a rarity or are they to be expected under appropriate circumstances.

In this paper, we construct a family of holographic models dual to anisotropic states in a strongly coupled gauge theory. Our construction is based on the near-horizon limit of a stack of $N_c$ D3-branes intersecting along $2+1$ dimensions with $N_f$ D5-branes. We take the Veneziano limit where $N_c\to \infty$ and $N_f/N_c$ remains fixed. On the gravity side this is realized by considering the backreaction of D5-branes in the geometry sourced by the D3-branes. The D5-branes are smeared along the transverse directions parallel to the D3-branes, in such a way that the resulting solution is homogeneous but anisotropic along one of the spatial directions of the field theory dual. Configurations of this type were previously constructed and studied in \cite{Penin:2017lqt,Jokela:2019tsb,Gran:2019djz}. The main novelty in this work is that we allow the density of D5-branes to go to zero at the asymptotic boundary of space.  Similar supergravity solutions have been constructed in \cite{Conde:2011ab,Conde:2011aa} to study the Higgsing and Seiberg dualities of cascading theories and their relations with the tumbling phenomena in theories of extended technicolor.  From the point of view of the field theory dual this means that instead of modifying the action by adding additional degrees of freedom localized on the $(2+1)$-dimensional defects, the anisotropy is produced spontaneously. This is similar to the anisotropic $p$-wave superfluids, except that the anisotropy is present even at zero density. In the case at hand, the operator that acquires an expectation value is a three-form and has conformal dimension $\Delta=3$. In four dimensions it is related by Hodge duality to an axial vector field, thus parity is unbroken. The operator is in a non-trivial representation of the $R$-symmetry group, which is then also spontaneously broken. This is reflected in the dual geometry as a deformation of the internal space.

The configurations we find are realized at vanishing temperature and density. They are also supersymmetric, thus stability is guaranteed. As far as we are aware there are no other examples in the literature with these characteristics. It should be mentioned that although we based our construction on a string theory setup, we have not shown that the D5-brane density we use can actually be obtained from the smearing of localized D5-branes, so our construction is phenomenological in this sense. One may ask the question of how a state of this type might be reached, a possibility is that the system was put under the action of an external force that induced the anisotropy and, when the force was turned off, the system remained in an anisotropic state. This would be analogous to what happens to a lump of iron when it is put in the presence of a magnet. The iron is magnetized and remains in this state even after the magnet is removed.

In the UV, the field theory flows to an isotropic fixed point, the well-studied ${\cal{N}}=4$ Yang-Mills in $(3+1)$ dimensions. In the IR, we find two distinct behaviors depending on the density of D5-branes close to the origin of the bulk. If the density falls fast enough, the theory follows a boomerang flow and goes to an isotropic fixed point similar to the one in the UV.  Our analysis thus indicates that boomerang flows appear quite generically in holographic duals if the deformation is irrelevant enough in the IR. If the density goes to zero more slowly, or goes to a constant, the IR is Lifshitz-like: there is an associated scaling symmetry of the anisotropic spatial direction. In order to characterize the flow we study the evolution of anisotropy and use different proposals for $c$-functions, none of which turn out to be monotonic. However, some quantities have lower values in the IR than in the UV, so a weaker version of the $c$-theorem might exist for anisotropic systems.

The paper is organized as follows. We begin in Sec.~\ref{sec:setup} by laying out the ten-dimensional background geometry and pay special attention to both the UV and IR regimes. We also discuss the field theory interpretation of our supergravity solution by first consistently reducing the geometry to five dimensions and then identifying the operator in the UV conformal theory that is responsible for the breaking of the isotropic symmetry spontaneously. We then continue in Sec.~\ref{sec:properties} to analyze the solution. We define an effective Lifshitz exponent at any energy scale. We also discuss different definitions for the $c$-functions via null congruences and via entanglement entropies. Sec.~\ref{sec:discussion} contains our final thoughts and future directions that we aim to study.

\section{The supergravity solution}\label{sec:setup}

In this section we will briefly discuss the ten-dimensional background geometry that we have constructed. We will also outline the user-friendly effective action for five-dimensional bulk geometry which can be directly adopted in various applications.

\subsection{Background geometry}

Let us consider the following array of $N_c$ D3-branes and $N_f$ D5-branes:
\be
\begin{array}{cccccccccccl}
 &0&1&2&3& 4& 5&6 &7&8&9 &  \\
(N_c)\,\,D3: &\times & \times &\times &\times & - &- & -&- &- &- &      \\
(N_f)\,\,D5: &\times &\times&\times&-&\times&\times&\times&-&-&- &
\end{array}
\label{D3D5intersection}
\ee
In (\ref{D3D5intersection}) the D3-branes are color branes which generate an $AdS_5\times {\mathbb S}^5$ space dual to  ${\cal N}=4$ super Yang-Mills (SYM), a gauge theory in four spacetime dimensions. The D5-branes create a codimension one defect which deforms anisotropically the $(3+1)$-dimensional theory. This deformation is reflected in the ten-dimensional metric when the backreaction of the D5-branes is taken into account. To find these backreacted geometries we will follow the smearing approach (see \cite{Nunez:2010sf} for a review) and will homogeneously distribute the D5-branes in such a way that a residual amount of supersymmetry is preserved.  The general form of the smeared type IIB backgrounds corresponding to the D3-D5 array in (\ref{D3D5intersection}) was found in \cite{Conde:2016hbg} (see also \cite{Penin:2017lqt,Jokela:2019tsb}).  To write the deformed metric, let us represent the five-sphere ${\mathbb S}^5$ as a $U(1)$ bundle over ${\mathbb C}{\mathbb P}^2$. The ten-dimensional backreacted metric can then be written as
\bea
ds^2_{10} &  = & h^{-{1\over 2}}\,\big[-(dx^0)^2+(dx^1)^2+(dx^2)^2\,+\,e^{-2\phi}\,(dx^3)^2\big]\nonumber\\
          &    & +h^{{1\over 2}}\,\Big[\zeta^2 e^{-2f}\,d\zeta^2\,+\,\zeta^2\,ds^2_{{\mathbb C}{\mathbb P}^2}\,+\,e^{2f}\,(d\tau+A)^2\Big]\ ,\label{metric_ansatz_zeta}
\eea
where $\phi$ is the dilaton of type IIB supergravity, $h$ is the warp factor, and $f$ is the squashing function of the internal space. These functions are assumed to depend only on the radial holographic coordinate $\zeta$; boundary is at $\zeta=\infty$ and the origin of spacetime is at $\zeta =0$. Moreover, $A$ is a one-form on ${\mathbb C}{\mathbb P}^2$ inherent to the non-trivial $U(1)$ bundle.  The preservation of two supercharges for our Ansatz leads to a series of first-order differential equations for the functions in (\ref{metric_ansatz_zeta}).  These equations can be combined and reduced to single second-order equation for a master function $W(\zeta)$ \cite{Conde:2016hbg,Jokela:2019tsb}, in terms of which $f$ and $\phi$ are given by
\be
e^{2f}\,=\,{6\,\zeta^2\,W\over 6\,W\,+\,\zeta\,{dW\over d\zeta}} \ , \ e^{-\phi}\,=\,W\,+\,{1\over 6}\,\zeta\,{d W\over d\zeta}\ .
\label{f_phi_W}
\ee
The warp factor $h$ can be written in terms of the following integral
\be
h(\zeta)\,=\,Q_c\,e^{-\phi(\zeta)}\,\int_{\zeta}^{\infty}{d\bar \zeta\over \bar\zeta^5\,W(\bar\zeta)}\ ,
\label{h_W}
\ee
where $Q_c$ is related to the number $N_c$ of D3-branes as follows
\be
Q_c\,=\,16\,\pi\,g_s\,\alpha'^{\,2}\,N_c\ .
\ee
The second-order differential equation satisfied by the master function $W$ is:
\be
 {d\over d\zeta}\Big(\zeta\,{d W\over d\zeta}\Big)\,+\,6\,{d W\over d\zeta} = -{6\,Q_f\,p(\zeta)\over \zeta^2\,\sqrt{W}}\ ,\label{Master_W}
\ee
where $Q_f$ is a constant proportional to the number $N_f$ of D5-branes and $p(\zeta)$ is a profile function which characterizes the distribution of D5-branes along the holographic direction $\zeta$.  The type IIB supergravity background is complemented with Ramond-Ramond three- and five- forms, whose explicit expressions are written for completeness in Appendix~\ref{Background_details}. They, apart from elucidating the field theory connection, do not play a significant role in the current paper.

The undeformed $AdS_5\times {\mathbb S}^5$ solution corresponds to taking $p=0$ and $W=1$. In this paper, we are interested in the case in which the geometry becomes $AdS_5\times {\mathbb S}^5$ only asymptotically in the UV and thus $W(\zeta)\to 1$ and $p(\zeta)\to 0$ in the region $\zeta\to\infty$. We will argue in Sec.~\ref{sec:fieldtheory} that we can achieve this by allowing a VEV for a three-form field, which then induces anisotropy at lower energy scales.  As shown in \cite{Conde:2016hbg}, the smeared D5-branes contribute to the energy density as $T_{00}^{D5}\propto 3p+e^{f}dp/d\zeta$.   Clearly, this expression is not positive definite in general when $p$ decreases with $\zeta$, as it happens for large $\zeta$.  However, in the asymptotic $AdS_5\times {\mathbb S}^5$ geometry $e^{f}\sim\zeta$ for large $\zeta$ and so the positive energy condition is tantamount to demanding that $p(\zeta)$ should decrease as $p\sim\zeta^{-3}$ or more slowly, if we want to have a positive energy density $T_{00}^{D5}$ in the UV, in such a way that we can interpret the solution as sourced by ordinary D5-branes with positive tension. We have succeeded in finding a two-parameter family of solutions fulfilling this requirement. These solutions are derived in detail in Appendix~\ref{Background_details}. Let us now illustrate that the above properties are satisfied by our solutions. The master function of these solutions reads as follows
\bea
W(\zeta) & = & 1\,+\,Q_f\,\,\Bigg[{1\over 4(\kappa\, \zeta)^4}\,F\Big({4\over m}, {3+n\over m}; {4+m\over m};- (\kappa\, \zeta)^{-m}\Big)\nonumber \\
& & +{(\kappa\,\zeta)^{n-1}\over 5+n} F\Big({5+n\over m}, {3+n\over m}; {5+m+n\over m};- (\kappa\, \zeta)^m\Big)\Bigg]\ ,\label{W_general_sol}
\eea
where $F$ are hypergeometric functions, $\kappa$ is a constant with units of mass and $n$ and $m$ are arbitrary non-negative dimensionless constants. However, we will later show that for physical considerations, we need to restrict the allowed domain for solutions (\ref{W_general_sol}) to
\be
  n\geq \frac{1}{3} \ , \ 4 > m > 0 \ .
\ee
\begin{figure}[ht]
\center
 \includegraphics[width=0.6\textwidth]{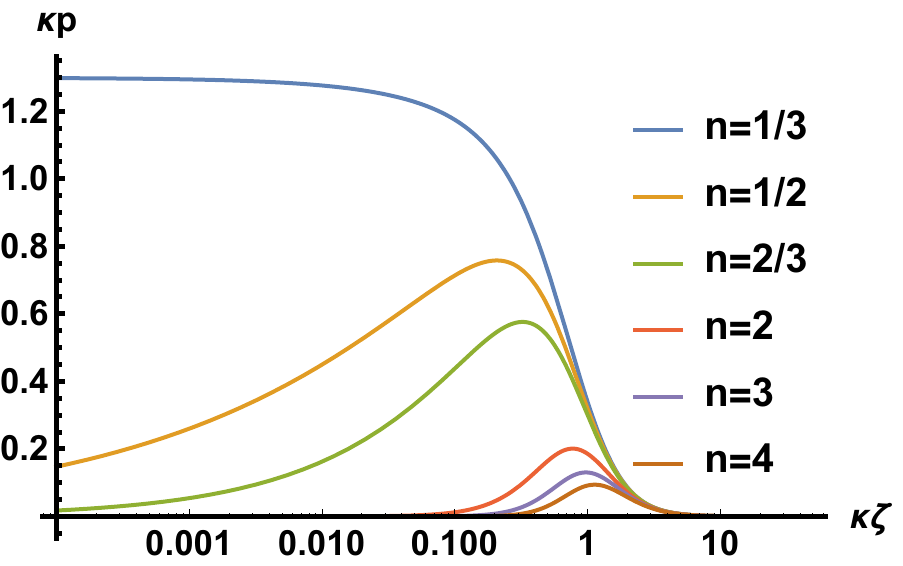}
  \caption{We illustrate the profile function (\ref{general_profile}) for various choices of $n$, while keeping $m=2$ and $Q_f=1$ fixed. The value of $n$ increases as gazing curves from top to bottom. The profile peaks at some $O(1)$ radial coordinate, which will be indicated in the coming figures with asterisk symbols.}
\label{fig:profile}
\end{figure}
Finally, the profile function corresponding to (\ref{W_general_sol}) is
\be\label{general_profile}
 \kappa\,p(\zeta) = \sqrt{W(\zeta)}{(\kappa\zeta)^n\over \big(1+(\kappa\zeta)^m\big)^{{n+3\over m}}}\ .
\ee
In Fig.~\ref{fig:profile} we have depicted the profile function for select values of $n$ to show that for all cases it vanishes rapidly enough at the UV, but in the IR it either vanishes ($n>1$) or goes to a constant ($n<1$). The profile has a global maximum at some intermediate energy scale. In the figures to follow we have indicated these global maxima by asterisks.

Given the master function we can construct all the functions of the supergravity solution; see Appendix~\ref{Background_details}. We have checked that the resulting geometry is free from curvature singularities.
In the following, let us focus on the asymptotic behaviors.

\subsubsection{UV regime}

The expansion in the UV region of the geometry $\zeta\to \infty$ is
\be
 W = 1+\frac{3Q_f }{4(\kappa \zeta)^4}+\ldots \ , \ \zeta\to\infty \ .
\ee
This indeed yields a sufficiently rapidly decreasing density $p(\zeta)\sim\zeta^{-3}$:
\be
 p = \frac{\kappa^{-1} }{(\kappa \zeta)^3}\left(1+\frac{3Q_f }{8(\kappa \zeta)^4}-\frac{n+3}{m(\kappa\zeta)^m}+\ldots \right) \ . 
\ee
Let us next show that the background is sourced by branes with positive tension. The behavior of the energy density of the D5-branes close to the boundary is
\be
T_{00}^{D5}\propto 3p+e^f p'\simeq \frac{\kappa^{-1}}{(\kappa \zeta)^3}\left[ -\frac{9 Q_f }{4(\kappa \zeta)^4}+\frac{n+3}{(\kappa\zeta)^m}\right] \ .
\ee
In order to remain positive asymptotically, the first term should decay faster than the second, which restricts $m<4$.\footnote{In the case $m=4$ we could still have $T_{00}^{D5}>0$ if the density of D5 branes is small enough $Q_f\leq \frac{4(n+3)}{9}$ (the subleading term is positive when the bound is saturated), however, we will not study this possibility.}. Assuming this condition holds, the expansions of the dilaton and the warp factors are
\be
 e^{-\phi}\simeq 1+\frac{Q_f }{4(\kappa \zeta)^4} \ , \ e^{2 f}\simeq \zeta^2\left( 1+\frac{Q_f }{2(\kappa \zeta)^4}\right) \ , \  h\simeq \frac{Q_c}{4\zeta^4} \ .
\ee
Therefore, the dilaton vanishes asymptotically and the geometry approaches $AdS_5\times {\mathbb S}^5$ with radius $\ruv^4 =Q_c/4$.

\subsubsection{IR regime}

Away from the UV region the metric becomes anisotropic, but it does not necessarily stay anisotropic indefinitely. The behavior in the IR $\zeta \to 0$ depends on the profile of the D5-brane density, in particular, on the value of the exponent $n$ in \eqref{general_profile}. We can distinguish two cases depending on whether $n>1$ or $n<1$, with a limiting case $n=1$ between the two. The master function has the following IR expansions, depending on the value of $n$,
\be\label{eq:irexpboom}
W\simeq  \left\{  \begin{array}{ll}
 w_{n,m}+\frac{6  Q_f }{(n+5)(1-n)}(\kappa\zeta)^{n-1} & , \ n>1 \\
 -Q_f \log (\kappa \zeta) & , \ n=1 \\ 
 \frac{6 Q_f }{(n+5)(1-n)}(\kappa \zeta)^{n-1} & , \ n<1 \ , \end{array} \right. 
\ee
where
\be\label{eq:wnm}
w_{n,m}\,=\,1+{\Gamma\Big({4\over m}\Big)\,\Gamma\Big({n-1\over m}\Big)\over m\,\Gamma\Big({3+n\over m}\Big)}\,Q_f \ .
\ee
From these expressions one can infer the expansion for the D5-brane density
\be
 p \simeq \left\{\begin{array}{ll}
 \kappa^{-1} \sqrt{w_{n,m}}(\kappa \zeta)^n  & , \ n>1 \\
 \kappa^{-1} \sqrt{\frac{6 Q_f }{n+5}}  (\kappa \zeta)(-\log(\kappa\zeta))^{1/2} & , \ n=1 \\
 \kappa^{-1}  \sqrt{\frac{6 Q_f }{(n+5)(1-n)}}(\kappa \zeta)^{\frac{3n-1}{2}} & , \ n<1 \ .\end{array}\right. \label{eq:pforn1over3}
\ee
The density at $\zeta=0$ remains finite as long as $n\geq 1/3$.  In the IR region $p$ is growing and positive for $n>1/3$, in which case it is guaranteed that $T_{00}^{D5}\geq 0$. The limiting case $n=1/3$ matches with the behavior of a constant density of massless defects constructed in \cite{Conde:2016hbg}.

The behavior of the metric is qualitatively different in the case $n>1$ and $n<1$. For $n>1$ the solution resembles the ``boomerang'' flow \cite{Donos:2017ljs},\footnote{Notice, however, that in \cite{Donos:2017ljs} translation invariance is explicitly broken.} in the sense that in the IR the geometry becomes isotropic again and approaches $AdS_5\times {\mathbb S}^5$ with the same radius as the UV geometry:
\bea
 e^{-\phi} & \simeq & w_{n,m}-\frac{ Q_f }{n-1}(\kappa\zeta)^{n-1}\simeq w_{n,m}\nonumber\\
 e^{2f} & \simeq & \zeta^2\left( 1+\frac{ Q_f }{w_{n,m}\,(n+5)}(\kappa\zeta)^{n-1}\right)\simeq \zeta^2 \nonumber\\
 h & = & \frac{Q_c}{4\zeta^4}\left(1+{\cal O}((\kappa \zeta)^{n-1})\right) \ .
\eea
The only difference between the UV and IR geometries is the magnitude of the dilaton, {\emph{i.e.}}, the coupling constant has flown, and that the length scale in the direction transverse to the D5-branes has been renormalized by a constant factor. In the $n=1$ case the metric deviates from the $AdS$ solution by logarithmic factors. From now on, we will not consider $n=1$ any further.

When $n<1$ the anisotropy along the spatial direction transverse to the D5-branes survives in the IR and the geometry becomes of Lifshitz-type. The expansion of the dilaton and warp factors of the metric is ($\zeta\to 0$):
\bea
e^{-\phi} & \simeq &  \frac{Q_f }{1-n}(\kappa \zeta)^{n-1}\nonumber\\
e^{2f} & \simeq &  \frac{6}{n+5}\zeta^2\nonumber\\
h & = & \frac{n+5}{6(n+3)}\frac{ Q_c}{\zeta^4}\left(1+{\cal O}((\kappa \zeta)^{1-n})\right) \ .
\eea
Let us write the ten-dimensional IR metric as:
\be
 ds^2_{IR} = ds^2_5\,+\,d\hat s^2_5 \ .
\ee
After a convenient rescaling of the Minkowski coordinates, the non-compact part of the metric can be written as:
\be
 ds^2_5 = {\zeta^2\over R^2}\,\Big[-(dx^0)^2+(dx^1)^2+(dx^2)^2+(\mu\zeta)^{2(n-1)}\,(dx^3)^2\Big]+{R^2\over \zeta^2}\,d\zeta^2\ ,
\ee
where $\mu^{n-1}=Q_f \kappa^{n-1}/(1-n)$ and the radius $R$ is given by:
\be
 R^4 = \left(\frac{n+5}{6}\right)^3 \frac{Q_c}{n+3} = \frac{4}{n+3}  \left(\frac{n+5}{6}\right)^3\ruv^4 \ .
\ee
The compact part of the metric is a squashed version of ${\mathbb S}^5$, namely:
\be
d\hat s_5^2\,=\,{\hat R}^2\,\Big[ds^2_{{\mathbb C}{\mathbb P}^2}\,+\,{6\over n+5}\,(d\tau+A)^2\Big]\ , 
\ee
where the radius $\hat R$ is related to  $R$  as:
\be
\hat R^4 = \left({6\over n+5}\right)^2\,R^4 = {n+5\over 6(n+3)}\,Q_c= {n+5\over 6}{4\over n+3}\, \ruv^4 \ .
\ee
Notice that the non-compact part of the metric is invariant under the following anisotropic scale transformations:
\be
\zeta\,\to\, \zeta/\Lambda\ ,
\qquad\qquad
x^{0,1,2}\,\to \,\Lambda\, x^{0,1,2}\ ,
\qquad\qquad
x^{3}\,\to\,\Lambda^{n}\,x^{3}\ ,
\label{scaling_of_coordinates}
\ee
where $\Lambda$ is an arbitrary positive constant. This means that, effectively, the $x^3$ direction has an anomalous scaling dimension.  In canonical convention, with a general Lifshitz-like anisotropic scaling, the coordinates transforming as in (\ref{scaling_of_coordinates}), with $x^3\to \Lambda^{1\over z}\,x^3$, the dynamical exponent $z$ is a measure of the degree of anisotropy associated with this coordinate direction. Thus, in our model
\be\label{eq:zIR}
z = {1\over n} \ , \ n<1 \ .
\ee
Notice also that the dilaton transforms as $e^{\phi}\to\Lambda^{n-1}\,e^{\phi}$. In Sec.~\ref{sec:zeff} we will discuss the running of the dynamical exponent in more detail.

\subsection{Field theory interpretation}\label{sec:fieldtheory}

To complete this section we give a field theory interpretation of the solutions presented above. With this purpose it is convenient to formulate our backgrounds as solutions of a five-dimensional gravity theory. This reduced theory was obtained in  \cite{Penin:2017lqt} for the case of massless flavors, in which case the profile $p$ is constant everywhere; recall that this is also the IR limiting case for $n=1/3$ (\ref{eq:pforn1over3}). Here we will outline the generalization of the reduction to a non-trivial profile function (details are  given in Appendix~\ref{Background_details}). The reduction Ansatz for the metric is:
\be
 ds_{10}^2 = e^{{10\over 3}\,\gamma}\,g_{pq}\,dz^p\,dz^{q}\,+\,e^{-2(\gamma+\lambda)}\,ds^2_{{\mathbb C}{\mathbb P}^2}+e^{2(4\lambda-\gamma)}\,(d\tau+A)^2\ ,\label{10d_5d_metric_ansatz}
\ee
where $g_{pq}=g_{pq}(z)$ is a 5d metric and the scalar fields $\lambda$ and $\gamma$ depend on the 5d coordinates $z^p\,=\,(x^0, x^1,x^2,x^3, \zeta)$. As argued in \cite{Penin:2017lqt} the reduced theory has smeared codimension one branes and a gauge field strength ${\cal F}_4$, which originates from the reduction of the RR three-form. The reduced gravity action can be written in terms of these fields and the profile function $p(\zeta)$ (see Appendix~\ref{Background_details}). For the purposes of this section it is enough to consider the action of the gauge field  ${\cal F}_4$ which, up to a global constant factor,  takes the form:
\be\label{5d_action_gauge}
 S_{5d}^{gauge} =-{1\over 2\,\cdot \,4!}\,\int d^5z\, \sqrt{-g_5}\,\,e^{-4\gamma-4\lambda-\phi}\,( {\cal F}_4)^2 + \int {\cal C}_3\wedge \Sigma_2\ ,
\ee
where ${\cal C}_3$ is the three-form potential for  ${\cal F}_4=d\,{\cal C}_3$. The second term in (\ref{5d_action_gauge}) is a Wess-Zumino term, which depends on a smearing form $\Sigma_2$. In the reduced theory, $\Sigma_2$ encodes the distribution of the D5-brane charge.  The equation of motion for ${\cal F}_4$ is a standard Maxwell equation with a source,
\be\label{eom_F4_form}
 d\Big(e^{-4\gamma-4\lambda-\phi}\,*\,{\cal F}_4\Big) = -\Sigma_2\ .
\ee
In our solutions ${\cal F}_4$ can be written in terms of the profile and the dilaton as
\be\label{eq:F4}
 {\cal F}_4 = \sqrt{2}\,Q_f\,\zeta\,p(\zeta)\,e^{2\phi}\,d\zeta\wedge dx^0\wedge dx^1\wedge dx^2\ ,
\ee
whereas $\Sigma_2$ depends on the radial derivative of the profile and is given by
\be\label{Sigma_2}
 \Sigma_2 = \sqrt{2}\,Q_f\,p'(\zeta)\,d\zeta\wedge dx^3\ .
\ee

In the solutions we have constructed the distribution of five-brane charge goes to zero at the asymptotic boundary, so it does not change the UV field theory, which is still the dual to the theory living on the color three-branes, ${\cal N}=4$ SYM. There is nevertheless an RG flow that should be triggered by the expectation value of some operator. In \cite{Penin:2017lqt} it was shown that in the truncation to five-dimensions there is a background three-form potential that is proportional to the volume form of the five branes along the field theory directions. We expect that the operator acquiring an expectation value is the dual to this field. Consequently, if the five-brane distribution would be non-zero at the boundary we expect that the dual field theory is modified by introducing a non-zero coupling for the operator dual to the three-form.

The dual operator should be a three-form operator of conformal dimension $\Delta=3$, since the bulk three-form potential is massless. One should also remember that the three-form originates from a ten-dimensional Ramond-Ramond form that has non-zero components along the internal space. Those components break the isometries of the would-be $S^5$, thus the dual operator should break the $R$-symmetry of ${\cal N}=4$ SYM in the same way. Furthermore, as the original five-brane defect configuration on which the smeared distributions are based are parity 
invariant \cite{DeWolfe:2001pq}, the dual operator should preserve the same discrete symmetry as well.  A candidate Hermitian operator fulfilling these conditions can be constructed with the Majorana gaugino fields $\psi$,
\be
{\cal V}^{\mu\nu\rho}_a=-i{\rm Tr}\,\left(\overline{\psi} \gamma^{\mu\nu\rho} H_a \psi  \right) \ ,
\ee
where the trace is over the gauge group, $\gamma^{\mu\nu\rho}=\gamma^{[\mu}\gamma^\nu \gamma^{\rho]}$ is the completely antisymmetric product of three Dirac matrices, and $H_a$ is a Hermitian generator of the ${\cal N}=4$ SYM $R$-symmetry group $SO(6)\cong SU(4)$  in the $\mathbf{4}$ representation (corresponding to the gauginos). The components of the three-form that are sourced by a density of five-branes are the ones matching ${\cal V}_a^{012}$. 

In four spacetime dimensions the product of three gamma functions satisfies the special relation
\be
 \gamma^{\mu\nu\rho}=-i\epsilon^{\mu\nu\rho\sigma}\gamma_\sigma \gamma_5 \ .
\ee
The three-form operator is then the Hodge dual of an axial current
\be
 {\cal V}_a^{\mu\nu\rho}=-\epsilon^{\mu\nu\rho\sigma}{\rm Tr}\,\left(\overline{\psi}  \gamma_\sigma \gamma_5 H_a \psi  \right) \ .
\ee
More precisely, the ${\cal V}_a^{012}$ component is equal to an axial current in the direction transverse to the five-brane volume
\be
 {\cal V}_a^{012}={\rm Tr}\,\left(\overline{\psi} \gamma^3 \gamma_5 H_a \psi  \right) \ .
\ee

Assuming ${\cal V}_a^{\mu\nu\rho}$ is the correct identification for the dual operator to the three-form, we can compute its expectation value following the usual procedure of evaluating the on-shell gravitational action and taking a variation with respect the boundary values, the asymptotic boundary being at $\zeta\to \infty$ (UV). However, we should proceed with caution in order to identify the coupling of the dual operator correctly. The UV expansion of the fields ($\zeta\to \infty$) was given in the previous subsection. The metric approaches $AdS_5$ and it is easy to check that the dilaton $\phi$ and the scalar fields $\gamma$, $\lambda$ defined in \eqref{eq:5dscalars} all go to zero. In the case where the dual theory has $(2+1)$-dimensional defects smeared in the transverse directions, the density of D5-branes becomes constant at the asymptotic boundary  $p(\zeta)\simeq p_0$. The expansion of the four-form potential \eqref{eq:F4} is
\be
 F_4\simeq \,\sqrt{2} Q_f\,p_0\, \zeta \,d\zeta\wedge dx^0\wedge dx^1\wedge dx^2\ .
\ee
Therefore, the non-zero components of the three-form potential have the asymptotic expansion
\be
 C_{012}\simeq \frac{Q_f}{\sqrt{2}}\left( p_0\zeta^2+v_0\right) \ .
\ee
The two terms with coefficients proportional to $p_0$ and $v_0$ correspond to the leading and subleading solutions for a massless three form in $AdS_5$, respectively. If $p(\zeta)\to 0$ sufficiently fast at the boundary, as it is the case in the configuration we study, then only the term proportional to $v_0$ is present (plus subsubleading corrections). We will now show that this term corresponds to an expectation value.

The variation of the on-shell gravity action \eqref{5d_action_gauge} will give a boundary contribution
\be
 \delta S_{5d}^{on-shell} =-{1\over 6}\,\lim_{\zeta \to \infty} \int d^4x \, \sqrt{-g_5}\,\,e^{-4\gamma-4\lambda-\phi}\,F_4^{\zeta \mu\nu\rho} \delta C_{3\, \mu\nu\rho} \ .
\ee
Then,
\be
 \delta S_{5d}^{on-shell} \simeq  Q_f^2\ruv V_4\,\lim_{\zeta \to \infty} \, \left( p_0 \delta p_0\zeta^2+p_0 \delta v_0\right) \ ,
\ee
where $V_4$ is the regulated volume along the field theory directions. As usual, the on-shell action is divergent. In order to remove the divergence we need to add a boundary counterterm. This can be achieved by including a mass term for the three-form
\be
 S_{c.t.} = \lim_{\zeta\to \infty} \frac{1}{6\ruv} \int d^4 x \sqrt{-h_4} C^{\mu\nu\rho}C_{\mu\nu\rho} \ ,
\ee
where $h_4$ is the determinant of the induced boundary metric $h_{\mu\nu} = \frac{\zeta^2}{\ruv^2} \eta_{\mu\nu}$ with which the indices are raised.
The variation of the counterterm gives
\be
 \delta S_{c.t.}=\lim_{\zeta\to \infty} \frac{2}{\ruv} \int d^4 x \sqrt{-h_4} C^{012}\delta C_{012}\simeq -Q_f^2 \ruv V_4 \lim_{\zeta\to \infty}  \,\frac{1}{\zeta^2}(p_0\zeta^2+v_0)(\delta p_0 \zeta^2+\delta v_0) \ .
\ee
The sum of the variations of the on-shell action plus the boundary term is finite
\be
 \delta S_{5d}^{on-shell}+\delta S_{c.t}=-Q_f^2 \ruv V_4\, v_0 \delta p_0 \ .
\ee
This shows that the variational principle is consistent with taking $p_0$ as the coupling to the dual $\Delta=3$ operator and consequently $v_0$ should be identified as the expectation value. This supports our expectation that the RG flows constructed with a five-brane density vanishing at the boundary are triggered by the expectation value of the operator dual to the three-form potential.

\section{Properties of the solutions}\label{sec:properties}

In this section we analyze different properties of our backgrounds. We start by measuring the degree of anisotropy of our metrics at different holographic scales. In particular, we aim to characterize the flow by measuring the number of degrees of freedom at different energy scales. Recall that the UV fixed point is that of pure glue $(3+1)$-dimensional ${\cal{N}}=4$ SYM. The number of degrees of freedom scale with the rank as $\sim N_c^2$, so as a reference we define the ``central charge'' in the UV as
\be\label{eq:cUV}
 \cuv = \frac{N_c^2}{4} \ .
\ee
We start by computing the effective dynamical exponent of anisotropy in the following subsection and discuss its behavior for different geometries that we have constructed. After this, we then device different functions that measure the number of degrees of freedom, constructed to match up with the UV value (\ref{eq:cUV}).

\subsection{The effective anisotropy exponent and refraction index}\label{sec:zeff}

Let us consider a metric of a holographic dual with four Minkowski directions $x^0$, $x^1$, $x^2$, and $x^3$, which is anisotropic along the third spatial direction $x^3$.  We define the effective anisotropic Lifshitz exponent $z_{eff}=z_{eff}(\zeta)$ as:
\be\label{z_eff_definition}
 {1\over z_{eff}(\zeta)}\equiv 1+\zeta\,{d\over d\zeta}\log \sqrt{\Big|{g_{x^3 x^3}\over g_{x^0 x^0}}\Big|} \ .
\ee
Clearly, $z_{eff}=1$ if the metric is isotropic. The deviations from unity signal anisotropy along the $x^3$ direction. In fact, the function (\ref{z_eff_definition}) determines how the anisotropy evolves as we change the holographic coordinate $\zeta$, {\emph{i.e.}}, as we vary the energy scale. It can be thought as the analogue of the beta function for the anisotropy. To illuminate the definition (\ref{z_eff_definition}), consider a geometry such that the Minkowski part of the metric has the following form:
\be\label{Lifshitz_metric}
 \zeta^2\,\big[\,-(dx^0)^2\,+\,(dx^1)^2\,+\,(dx^2)^2\,\big]\,+\,\zeta^{{2\over z}}\,(dx^3)^2 + \ldots\ ,
\ee
with $z$ being a constant exponent.  One readily finds that $z_{eff}$ is constant and equal to $z$ for the metric (\ref{Lifshitz_metric}). Moreover, this metric is invariant under the scaling transformation (\ref{scaling_of_coordinates}) with $n={1\over z}$.

\begin{figure}[ht]
\center
 \includegraphics[width=0.48\textwidth]{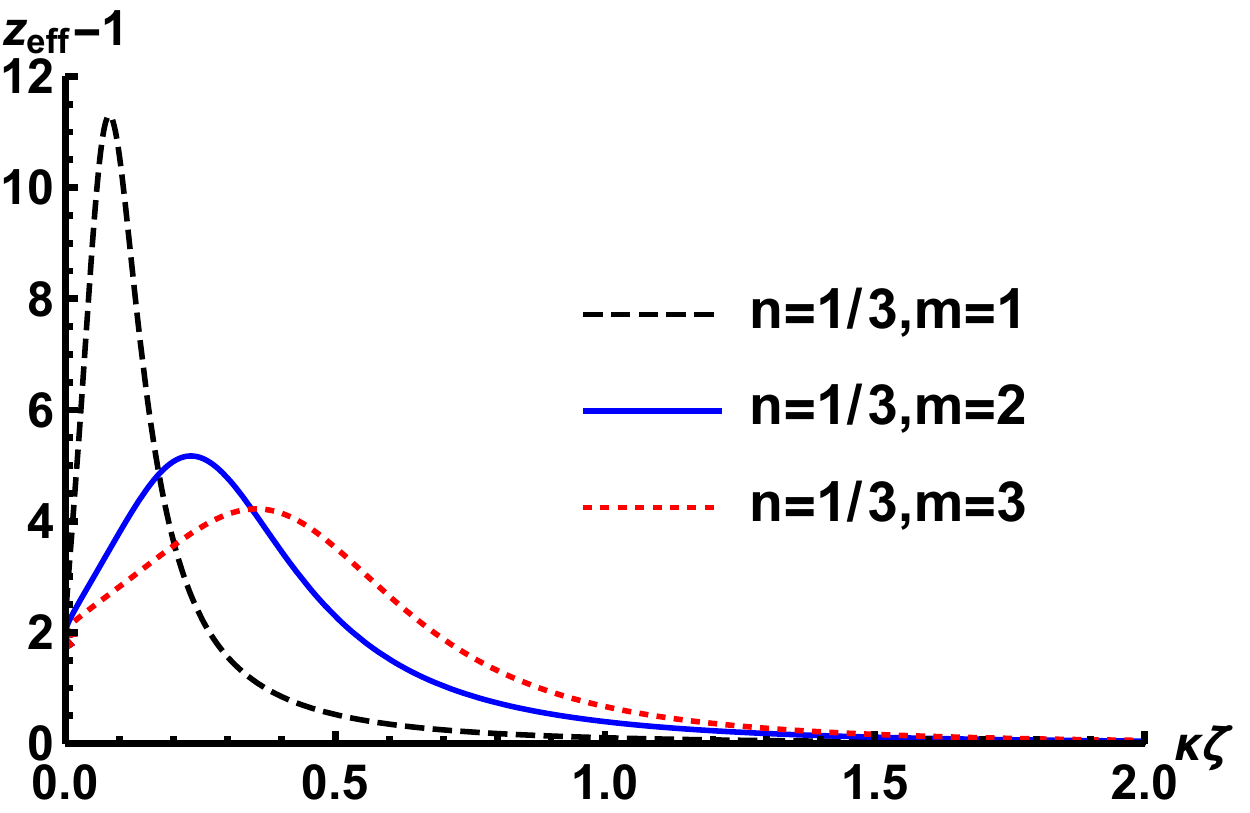}
 \includegraphics[width=0.48\textwidth]{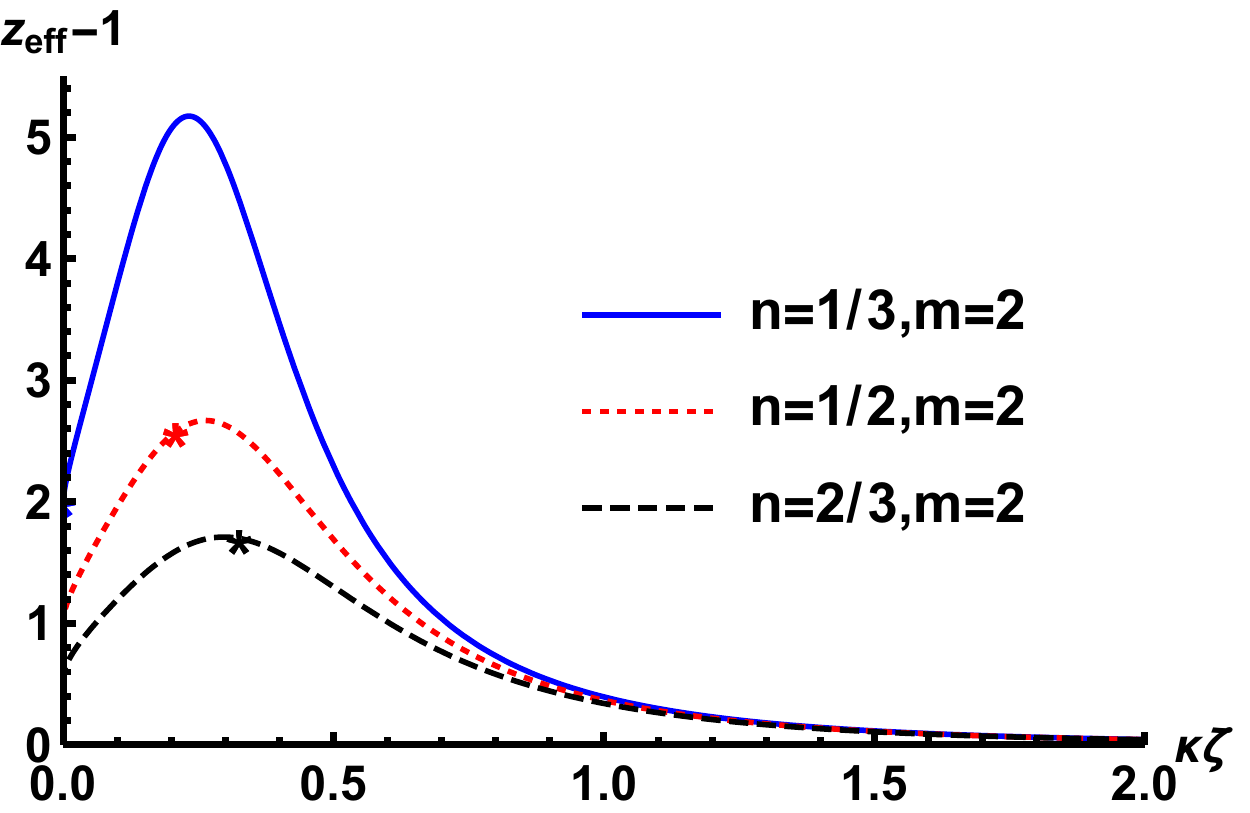}
  \caption{Plots of $z_{eff}(\zeta)-1$ for different anisotropic Lifshitz solutions for $Q_f=1$. Left: We depict $z_{eff}(\zeta)-1$ for $n=1/3$ and $m=1$ (dashed black), $m=2$ (blue), and $m=3$ (dotted red). Right: We plot the case $m=2$ with varying $n=1/3$ (blue), $n=1/2$ (dotted red), and $n=2/3$ (dashed black). The maximal exponent decreases for increasing $m$ or $n$.}
\label{Zeff_Lifshitz}
\end{figure}

\begin{figure}[ht]
\center
 \includegraphics[width=0.48\textwidth]{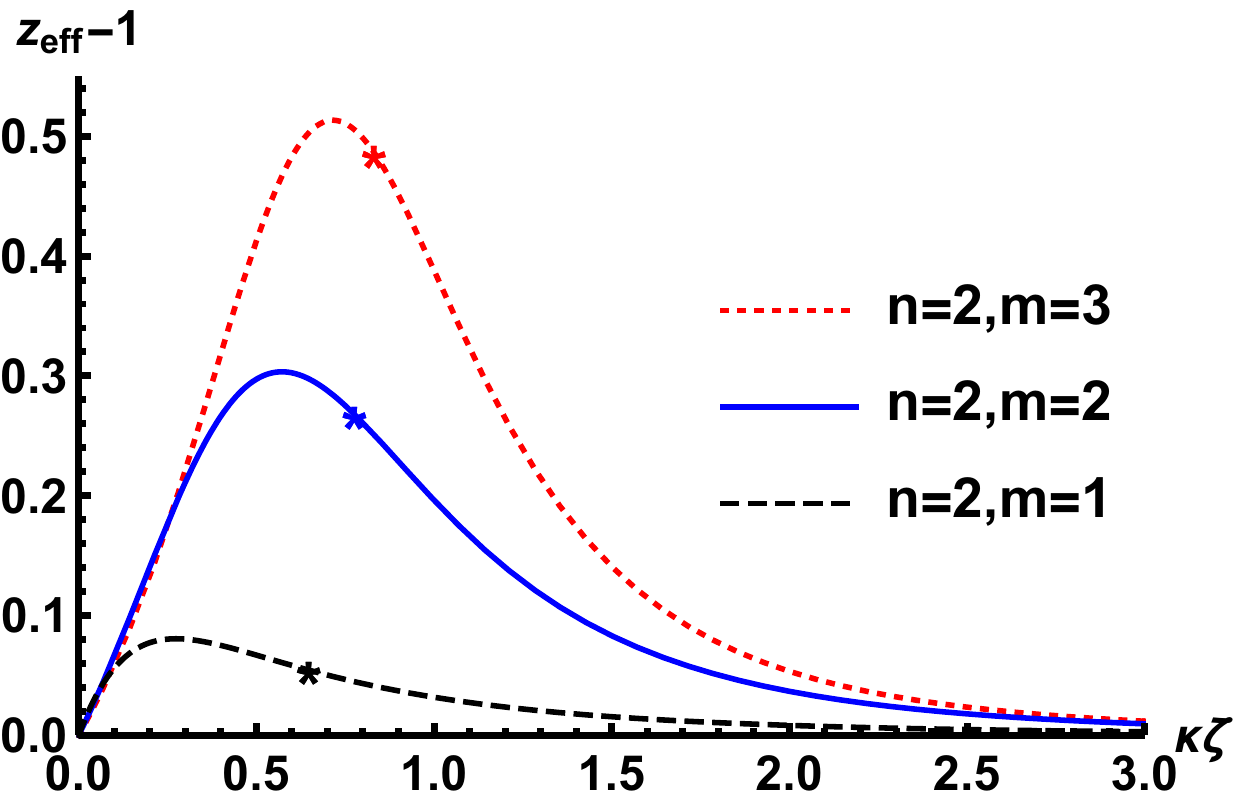}
 \includegraphics[width=0.48\textwidth]{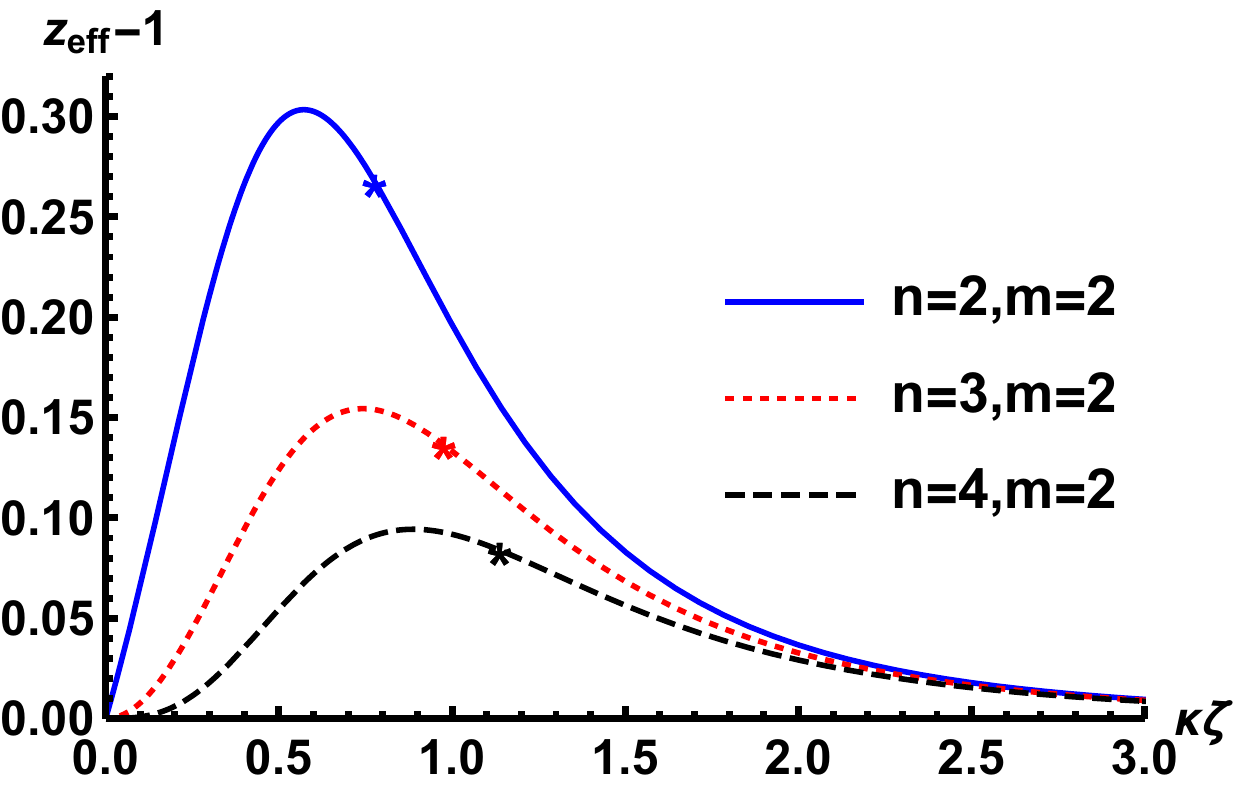}
  \caption{Relative effective anisotropy exponents $z_{eff}-1$ for boomerang flows for $Q_f=1$. Left: All the curves have the same $n=2$, while the different curves correspond to $m=1$ (dashed black), $m=2$ (blue), and $m=3$ (dotted red). Right: All the curves have the same $m=2$, but this time we vary $n=2$ (blue), $n=3$ (dotted red), and $n=4$ (dashed black). The maximal exponent decreases with increasing $n$, but increases with $m$.}
\label{Zeff_Boomerang}
\end{figure}

Let us now evaluate the function $z_{eff}(\zeta)$ for our anisotropic models. As:
\be
 \sqrt{\Big|{g_{x^3 x^3}\over g_{x^0 x^0}}\Big|}\,=\,e^{-\phi}\ ,
\ee
we can relate  $z_{eff}$ to the radial derivative of the dilaton:
\be\label{z_eff_p_phi_f}
 {1\over z_{eff}} =  1\,-\,\zeta\,{d\phi\over d\zeta}  = 1\,-\,Q_f\,p\,e^{{3\phi\over 2}-f}\ ,
\ee
where $p$ is the profile and $f$ is the squashing function of the metric (\ref{metric_ansatz_zeta}). It is possible to get a full analytic expression of $z_{eff}$ for the different values of $n$ and $m$. The derivation and the final result for this expression is presented in Appendix~\ref{app:Deg_ani}. Here we are content with only depicting the final result: $z_{eff}$ in Fig.~\ref{Zeff_Lifshitz} for anisotropic Lifshitz solutions and in Fig.~\ref{Zeff_Boomerang} for the boomerang solutions. Interestingly, for all values of $n$ and $m$ there is an intermediate region of $\zeta$ where  $z_{eff}(\zeta)$ has a maximum, {\emph{i.e.}}, the maximal anisotropy occurs at intermediate scales. 

It is, however, interesting to discuss the asymptotics. The behavior of $z_{eff}$ in the UV region $\zeta\to\infty$ reads
\be
 z_{eff}=1\,+\,Q_f\,(\kappa\zeta)^{-4}\,-\,{3+n\over m}\,Q_f\,(\kappa\zeta)^{-4-m}+\ldots\ , \ \zeta\to\infty \ .
\label{zeff_UV}
\ee
In all cases $z_{eff}(\zeta)\to 1$ as $\zeta\to\infty$, {\emph{i.e.}}, Poincar\'e invariance is retained in the UV. Notice also that the first UV anisotropic correction is independent of $n$ and $m$.  In the IR limit $\zeta\to 0$, 
\be\label{eq:zeff_IR}
 z_{eff} \simeq \left\{\begin{array}{ll}
  1+{Q_f\over w_{n,m}}\,(\kappa \zeta)^{n-1} & , \ n>1 \\
  {1\over n}-{(n-1)^2\,w_{n,m}\over n^2\,Q_f}\, (\kappa \zeta)^{1-n} & , \ n<1 \ .\end{array}\right. 
\ee
For the Lifshitz solutions one finds $z_{eff}(\zeta)\to 1/n$ as alluded to before in (\ref{eq:zIR}), while for the boomerang solutions one returns to the Poincar\'e invariant system. 

A quantity related to the anisotropic exponent is the refraction index $\mathfrak{n}=|g_{x^3x^3}/g_{x^0x^0}|=e^{-2\phi}$ \cite{Gubser:2009gp}. It was shown to be monotonically increasing towards the IR in the boomerang flows of \cite{Donos:2017sba}, and we find the same qualitative behavior in our configurations. The result follows from the equation that relates the dilaton to the master function \eqref{f_phi_W} and the master equation \eqref{Master_W}
\be
\frac{d}{d\zeta} \mathfrak{n} =-\frac{2 Q_f p}{\zeta^2\sqrt{W}}e^{-\phi} \leq 0 \ .
\ee
Note that it depends on the sign of the five-brane density, that we take to be positive as expected for physical D5-branes. If one follows a more bottom-up approach, and relaxes this condition,  the refraction index could also be engineered non-monotonic.

\subsection{Holographic Entanglement Entropy and $c$-functions}

According to the Wilsonian intuition, the number of degrees of freedom decreases effectively at large distances or low energies due to coarse graining. In two-dimensional field theories this was given a precise meaning through the definition of a $c$-function that equals the central charge of the CFT at UV and IR fixed points and that was shown to be monotonically decreasing along the RG flow; the celebrated $c$-theorem by Zamolodchikov  \cite{Zamolodchikov:1986gt}. A different version of the $c$-theorem based on the entanglement entropy was more recently derived by Casini and Huerta  \cite{Casini:2004bw,Casini:2012ei}. Using the subadditivity properties of entanglement entropy of a strip of length $\ell$, a $c$-function was defined as
\be\label{eq:cfunc2}
 c=3\ell\frac{\partial S_{EE}}{\partial \ell}\ .
\ee
This $c$-function is monotonically decreasing with $\ell$ and coincides with the central charge at the fixed points. Monotonic $c$-functions based on entanglement entropy have also been defined for field theories, {\emph{e.g.}}, in $2+1$  \cite{Casini:2012ei} dimensions. In $3+1$ dimensions there is a field theory proof of the $c$-theorem (the $a$-theorem)  \cite{Komargodski:2011vj}.  In theories with a holographic dual, a $c$-theorem exists for arbitrary dimensions, provided the null energy condition is satisfied in the bulk  \cite{Freedman:1999gp}. A generalization of \eqref{eq:cfunc2} to $D$ spacetime dimensions is suggested by a holographic computation  \cite{Myers:2010xs,Myers:2010tj},
\be\label{eq:cfuncD}
 c = \frac{1}{V_{D-2}}\beta_D \ell^{D-1}\frac{\partial S_{EE}}{\partial \ell}, \ \ \beta_D=\frac{1}{\sqrt{\pi}2^D \Gamma(D/2)}\left( \frac{\Gamma\left( \frac{1}{2(D-1)}\right)}{\Gamma\left( \frac{D}{2(D-1)}\right)}\right)^{D-1} \ ,
\ee
where $V_{D-2}$ is the area of the sides of the strip; it can be trivially regulated by implementing a periodic compactification in the spatial directions, for instance.
 
In all the aforementioned cases, the proof of the $c$-theorem utilizes Lorentz invariance in one way or another. There have been several attempts to find a monotonic $c$-function valid in holographic models with broken Lorentz invariance, with some partial success \cite{Liu:2012wf,Cremonini:2013ipa,Bea:2015fja,Chu:2019uoh,Ghasemi:2019xrl}.  As more recently shown in \cite{Chu:2019uoh}, for a theory with an anisotropic scaling symmetry
\be
t\to \Lambda t\ , \ x_i\to \Lambda^{n_1} x_i \ , \ y_j\to \Lambda^{n_2} y_j \ , \ i=1,\ldots,d_1 \ ,\ j=1,\ldots, d_2 \ ,
\ee
the entanglement entropy of an infinitely extended strip depends on the separation between the two sides $\ell$ with an exponent determined by the scaling exponents and the number of dimensions. For a strip separated along one of the $x_i$ directions,
\be
S_{EE}^{(x)}\sim -\frac{1}{\ell^{d_x}} \ ,
\ee
where $d_x=d_1-1+d_2\frac{n_2}{n_1}$. If the strip is separated along one of the $y_j$ directions, then
\be
S_{EE}^{(y)}\sim -\frac{1}{\ell^{d_y}} \ ,
\ee
where $d_y=d_2-1+d_1\frac{n_1}{n_2}$. These can be interpreted as the effective dimensions of the (hyper)planes on the sides of the strip divided by the effective dimension of the transverse direction. 

A clear question for the flows that we have constructed is whether a monotonic $c$-function can be defined through the entanglement entropy. Following the previous works we have mentioned, we will consider the entanglement entropy of strips with flat walls separated a distance $\ell$ along one of the spatial directions. According to the Ryu-Takayanagi (RT) prescription \cite{Ryu:2006bv,Ryu:2006ef}, the entanglement entropy is determined by a minimal codimension two surface in the gravity dual that lives on a fixed time slice and it is anchored at the $AdS$ boundary on the location of the sides of the strip. In the Einstein frame, the RT formula reads
\be
 S_{EE}=\frac{1}{4 G_{10}}\int d^8\sigma \sqrt{g_8} \ ,
\ee
where $g_8$ is the determinant of the induced metric on the surface and $G_{10}=8\pi^6$. In the anisotropic geometries we are studying, we have to distinguish between strips that are separated along the anisotropic direction, so the sides of the strip would be parallel to the defects described by D5-branes reaching the boundary of $AdS$,\footnote{In the type of geometries we are studying D5-branes do not reach the boundary, but the 5-brane charge distribution splits the spatial directions in the same way.} and strips separated along one of the other spatial directions, such that the sides of the strip will be crossing the defects. We will refer to the entanglement entropy (EE) of the first type as $S_{EE}^\parallel$ and of the second type as $S_{EE}^\perp$.  It should be noted that the results of \cite{Chu:2019uoh,Ghasemi:2019xrl} are obtained using domain wall coordinates and the conditions that 5d equations of motion impose on warp factors. The EE obtained by applying the RT prescription in the reduced 5d metric as defined in \eqref{10d_5d_metric_ansatz} is {\em different} from the EE obtained in the full 10d spacetime due to the non-trivial warp factors in the internal space in domain wall coordinates. Then, the results of \cite{Chu:2019uoh,Ghasemi:2019xrl} cannot be used directly for the EE we compute.

The calculation is standard (see Appendix~\ref{app:EE}) and gives the following expressions for the EE in the metric \eqref{metric_ansatz_zeta}
\bea
 S_{EE}^\parallel & =& \frac{\pi^3 V_2}{2G_{10}}\int_{\zeta_0}^{\zeta_\Lambda} d\zeta \frac{\zeta^5 h}{\sqrt{1-\frac{P^2e^{2\phi-2f}}{h\zeta^8}}} \nonumber \\
 S_{EE}^\perp & = & \frac{\pi^3 V_2}{2G_{10}}\int_{\zeta_0}^{\zeta_\Lambda} d\zeta \frac{\zeta^5 h e^{-\phi}}{\sqrt{1-\frac{P^2e^{2\phi-2f}}{h\zeta^8}}} .\label{eq:see}
\eea
Here $V_2$ is the area of the sides of the strip, which we consider finite via a periodic compactification of the spatial directions. There is the standard UV divergence from the integration along the radial direction: we have introduced a cutoff $\zeta_\Lambda$ in order to regularize it. The minimal surface that determines the EE consists of two sheets starting at the locations of the sides of the strip at the $AdS$ boundary, extending towards the bulk, and joining at the point $\zeta_0$, defined through an integration constant $P$:
\be
 \zeta_0^8 = P^2 h^{-1} e^{2\phi-2f}\Big|_{\zeta=\zeta_0} \ .
\ee
The EE depends implicitly on the separation between the two sides of the strip
\bea
 \ell_\parallel & = & 2P\int_{\zeta_0}^{\zeta_\Lambda} \frac{d\zeta}{\zeta^3} \frac{e^{2\phi-2f}}{\sqrt{1-\frac{P^2e^{2\phi-2f}}{h\zeta^8}}} \\
 \ell_\perp & = & 2P\int_{\zeta_0}^{\zeta_\Lambda} \frac{d\zeta}{\zeta^3}  \frac{e^{\phi-2f}}{\sqrt{1-\frac{P^2e^{2\phi-2f}}{h\zeta^8}}} .\label{eq:ell}
\eea
Using these expressions for the EE of the strips, we can mimic (\ref{eq:cfuncD}) by defining two possible ``$c$-functions'' as follows   
\be
 c_\parallel(\ell)=\frac{1}{V_2}C_\parallel(\ell)\frac{\partial S_{EE}^\parallel}{\partial \ell} \ , \ c_\perp(\ell)=\frac{1}{V_2}C_\perp(\ell)\frac{\partial S_{EE}^\perp}{\partial \ell} \ .
\ee
Desirable properties of the $c$-functions are that they become constants on scaling solutions and that they give the expected result in the UV. Concerning the second property, the UV expansion ($\ell\to 0$) of the EE is
\be
 S_{EE}^\parallel \simeq S_{EE}^\perp=\frac{\pi^3 V_2}{2G_{10}} \left(\frac{1}{2}\ruv^4\zeta_\Lambda^2-\frac{16c_0^3 \ruv^8}{\ell^2} \right)  \ ,
\ee
where $\ruv^4=Q_c/4$ and $c_0= \frac{\sqrt{\pi}\Gamma\left( \frac{2}{3}\right)}{2\Gamma\left(\frac{1}{6}\right)}$.
This means that for $\ell\to 0$,
\be\label{eq:CUV}
 C_\parallel(\ell) \simeq  C_\perp(\ell) \simeq \beta_4 \ell^3 \ .
\ee
The UV value of the $c$-function is fixed to the expected result (\ref{eq:cUV}), noting that $\beta_4=\frac{\pi}{128 c_0^3}$, yielding
\be
 \cuv = \lim_{\ell\to 0} c_\parallel(\ell)=\lim_{\ell\to 0} c_\perp(\ell)=\frac{\pi^4}{8} \frac{\ruv^8}{G_{10}}=\frac{N_c^2}{4} \ .
\ee

Before continuing to discuss the results for the entanglement entropies and the associated $c$-functions, let us make a brief comment. It turns out that if $Q_f$ is large enough, then there can be several competing minimal surfaces for large values of $\ell$. In the current paper we will choose to present results for $Q_f$ small enough to avoid addressing the issues related with phase transitions. 

 \subsubsection{$c$-functions in boomerang flows}

In order to describe the behavior of the solutions at a generic radial coordinate, we need a separate discussion depending on whether $n$ exceeds unity or not. 
Let us start with the boomerang flows, $n>1$. 

In the IR, the geometry becomes almost the same as in the UV, except for a finite rescaling of the anisotropic direction by the constant $w_{n,m}$ \eqref{eq:wnm}. The IR expansion $\ell \to \infty$ is
 \bea
 S_{EE}^\parallel & \simeq & \frac{\pi^3 V_2}{2G_{10}} \left(\frac{1}{2}\ruv^4\zeta_\Lambda^2-\frac{16c_0^3 \ruv^8 w_{n,m}^{-2}}{\ell^2} \right) \nonumber\\
 S_{EE}^\perp & \simeq & \frac{\pi^3 V_2}{2G_{10}} \left(\frac{1}{2}\ruv^4\zeta_\Lambda^2-\frac{16c_0^3 \ruv^8 w_{n,m}}{\ell^2}+{\rm constant} \right) \ . \label{eq:seeboom}
\eea
\begin{figure}[!ht]
\center
 \includegraphics[width=0.48\textwidth]{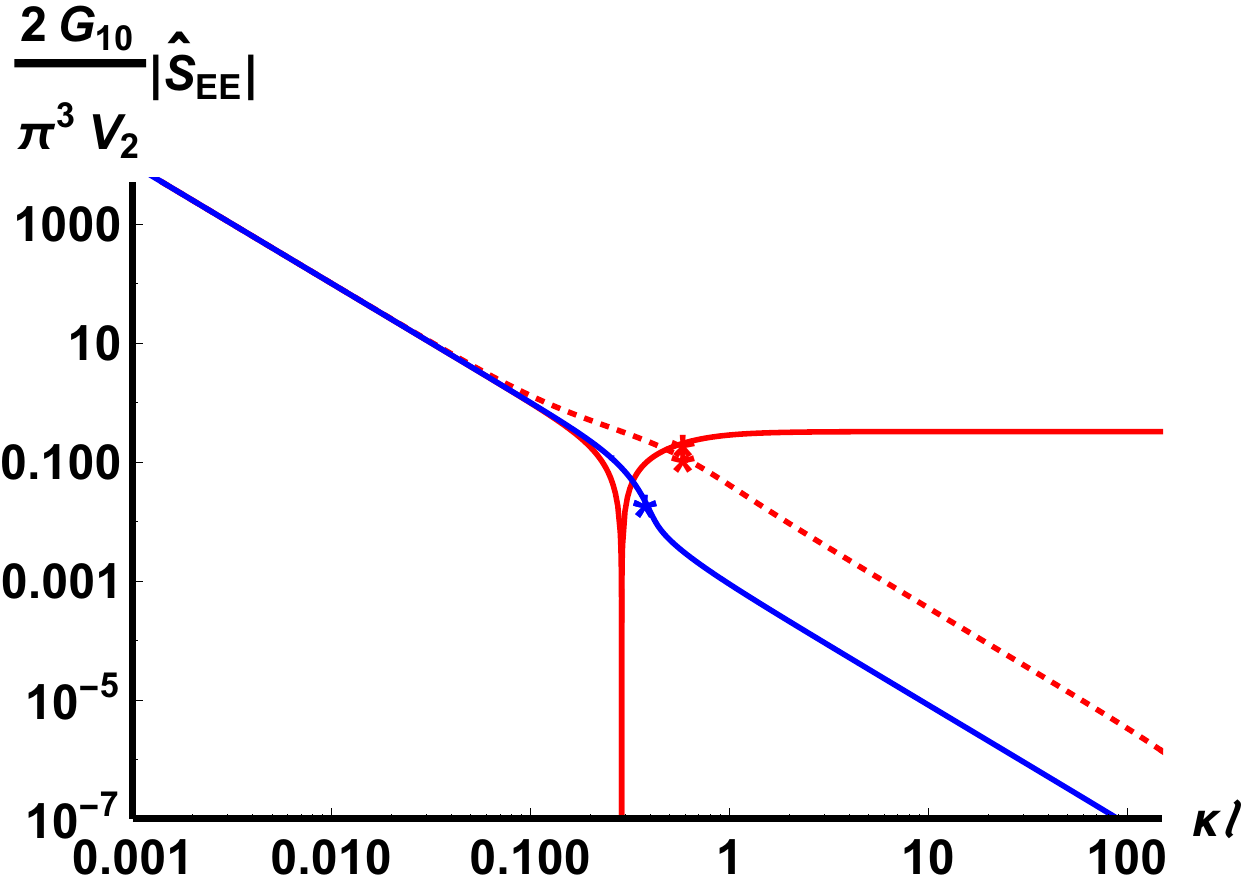}
 \includegraphics[width=0.48\textwidth]{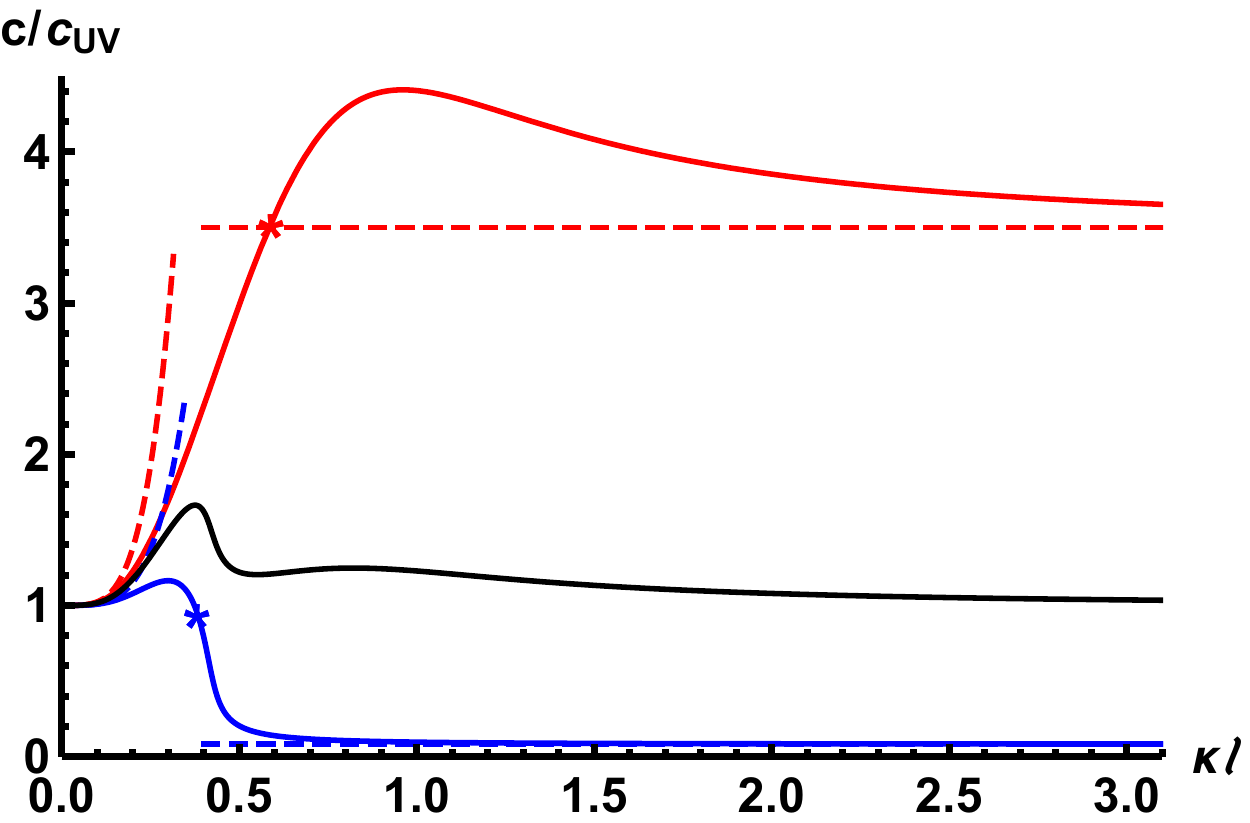}
\caption{Left: The (absolute values of the) regularized entanglement entropies for the in-plane ($\parallel$,blue) and for the off-plane ($\perp$,red) with the constant part in (\ref{eq:seeboom}) subtracted for the dotted curve to illustrate the $\ell^{-2}$ behaviors for the asymptotically narrow and wide slabs. Right: The $c$-functions for the boomerang flows. We chose as parameters $n=3,m=2$, but the results are qualitatively the same for other values. The flavor parameter we picked sizable $Q_f=10$ to pronounce the features. The solid curves are produced numerically, while the dashed curves follow from the asymptotic UV and IR analytics (\ref{eq:UVsubsub}) and (\ref{eq:boomerangIR}), respectively. The black curve is the average $c$-function defined in (\ref{eq:cavegboom}).}
\label{fig:boomerangcfunction}
\end{figure}
Since the scaling in the UV is the same as in the IR, a natural definition for the $c$-functions is in accord with that of UV CFT:
\be\label{eq:Cboomerang}
 C_\parallel=C_\perp  \equiv \beta_4 \ell^3 \ .
\ee
The IR value of the $c$-functions will be either larger or smaller than $\cuv$ depending on the orientation of the strip. We find
\be\label{eq:boomerangIR}
 \lim_{\ell \to \infty} c_\perp= w_{n,m} \cuv > \cuv > \lim_{\ell \to \infty} c_\parallel= w_{n,m}^{-2} \cuv \ .
\ee
Note that the following averaged $c$-function has the same values at the UV and IR, depicted in Fig.~\ref{fig:boomerangcfunction},
\be\label{eq:cavegboom}
\bar{c}=(c_\parallel c_\perp^2)^{1/3} \ .
\ee
The fact that degrees of freedom as measured with $c_\parallel$ dwindled, makes it a prospective candidate also for a monotonically decreasing $c$-function. However, we find that it is not monotonic, showing a global maximum away from the fixed points, around the intrinsic energy scale of the background, see Fig.~\ref{fig:boomerangcfunction}. 



\subsubsection{$c$-functions in flows with anisotropic IR}

Let us now discuss the flows with Lifshitz scaling in the IR. First, recall that the UV behavior does not change for these flows, the behavior of the $c$-functions in the UV, $\ell\to 0$, is as in \eqref{eq:CUV}.  The IR scalings along the $(x^1,x^2,x^3)$ directions can be taken to be $n_1=n_2=1$, $n_3=n<1$. We then expect the dependence of the EE with the width of the strip to be
\bea
 S_{EE}^{\parallel} & \sim & -\frac{1}{\ell^{\frac{n_1+n_2}{n_3}}}=-\frac{1}{\ell^{2/n}} \label{eq:inIR} \\
 S_{EE}^{\perp} & \sim & -\frac{1}{\ell^{\frac{n_{1,2}+n_3}{n_{2,1}}}}=-\frac{1}{\ell^{n+1}} \label{eq:offIR}\ .
\eea

\begin{figure}[!ht]
\center
 \includegraphics[width=0.48\textwidth]{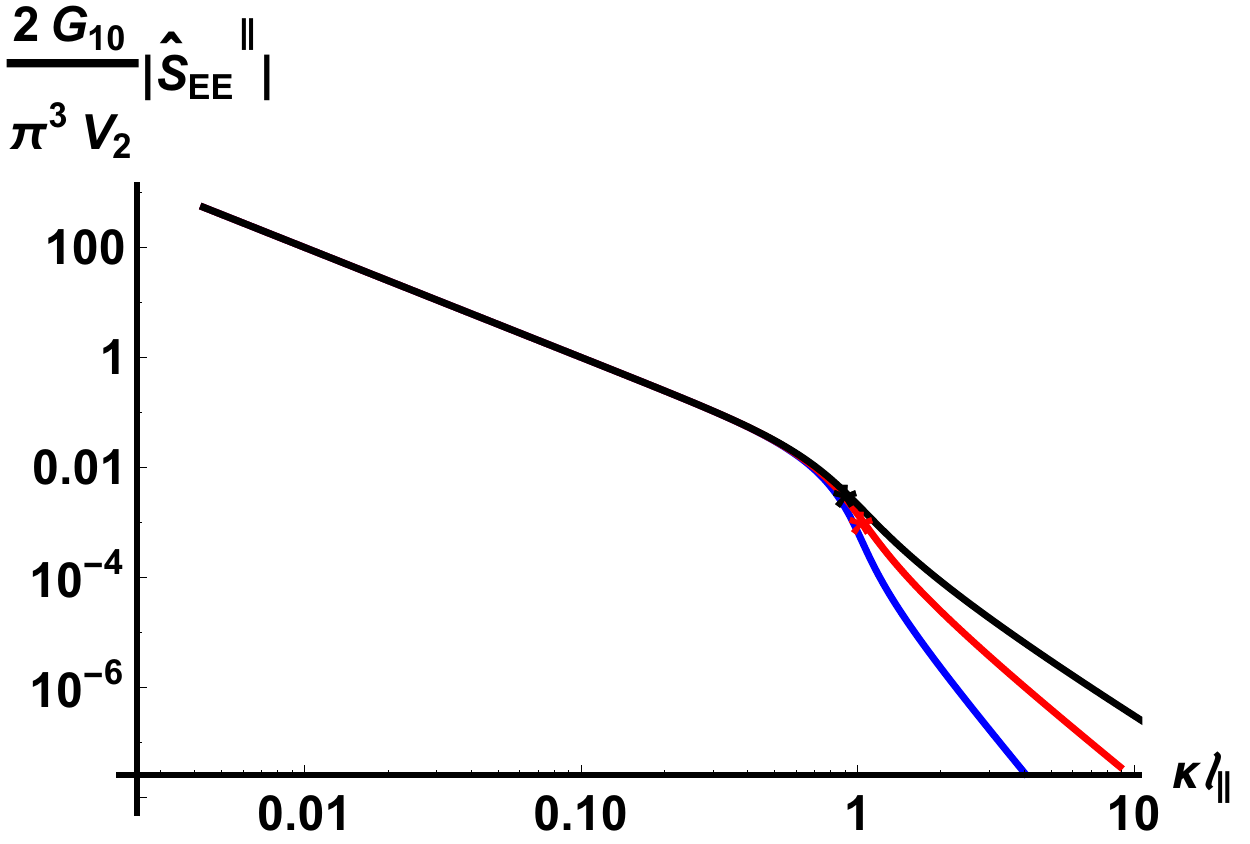}
 \includegraphics[width=0.48\textwidth]{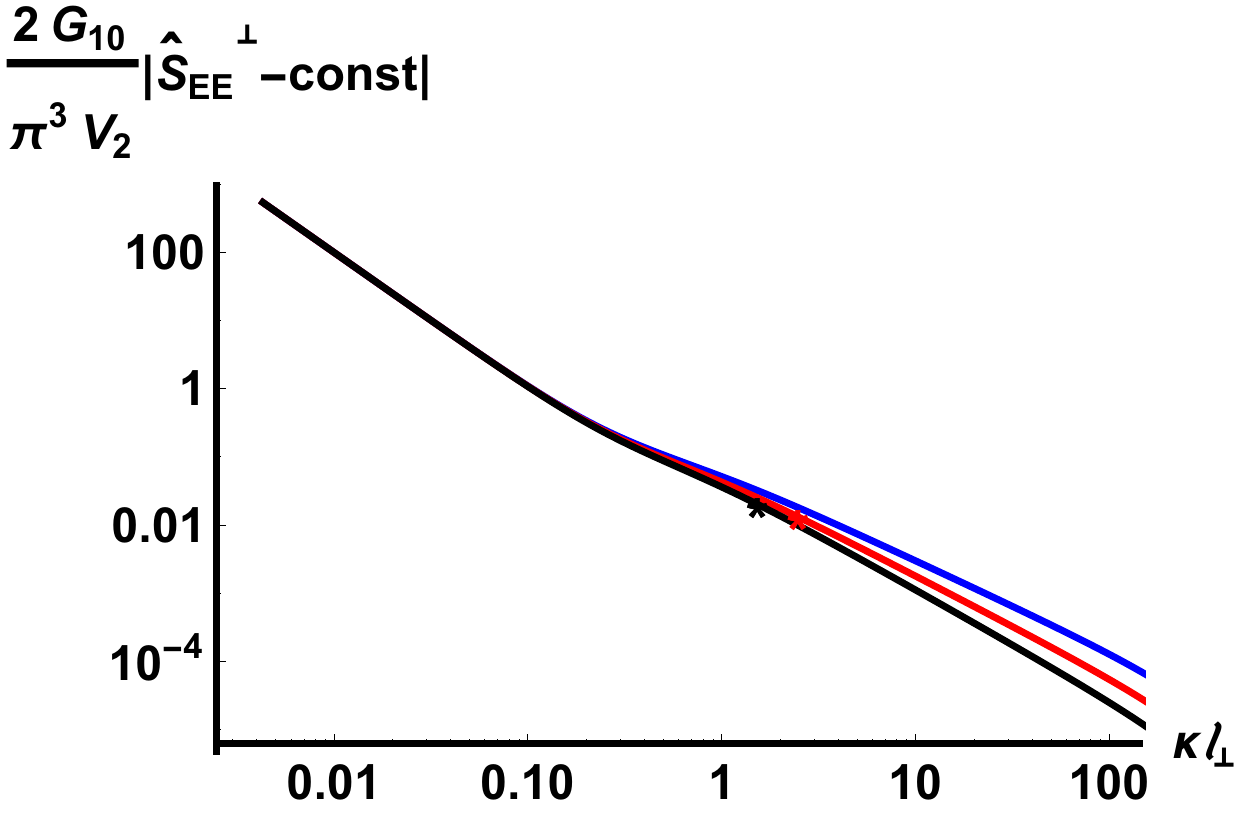}
  \caption{The regularized entanglement entropies for the anisotropic Lifshitz solutions with $m=2$ and $n=1/3$ (blue), $n=1/2$ (red), and $n=2/3$ (black) and $Q_f=1$. Left: The in-plane case. The slopes are -2 in the UV and $-2/n$ is the IR conforming with (\ref{eq:inIR}). Right: The off-plane case. The slopes are -2 in the UV and $-(n+1)$ in the IR conforming with (\ref{eq:offIR}).}
\label{fig:SEEaniso}
\end{figure}
Compared to the UV scaling, we see that the effective dimensions satisfy
\be
\frac{2}{n}=d_\parallel > d_{UV}=2 > d_\perp =n+1 \ .
\ee
Indeed, we find, for $\ell\to \infty$ (details are in Appendix~\ref{app:EE}), 
\be
 \begin{split}
&S_{EE}^\parallel\simeq \frac{\pi^3 V_2}{2G_{10}} \left(\frac{1}{2}\ruv^4\zeta_\Lambda^2-\frac{R^6A_\parallel B_\parallel^{2/n}}{(\mu R)^2}\left(\frac{\mu R^2}{\ell}\right)^{2/n} \right) \\
&S_{EE}^\perp\simeq \frac{\pi^3 V_2}{2G_{10}} \left(\frac{1}{2}\ruv^4\zeta_\Lambda^2-\frac{R^6A_\perp B_\perp^{n+1}}{(\mu R)^2}\left(\frac{\mu R^2}{\ell}\right)^{n+1}+{\rm constant} \right)\ .\label{eq:seescale}
\end{split}
\ee
Note that
\be
R^4=\frac{4}{n+3}\lambda_n^6 \ruv^4\ ,\ \lambda_n=\sqrt{\frac{n+5}{6}} \ .
\ee
Let us define 
\be\label{eq:lncn}
c_{n-1} ^\parallel=\frac{\sqrt{\pi} \Gamma\left(\frac{n+1}{n+2} \right)}{2\Gamma\left(\frac{n}{2(n+2)}\right)},\ \  c_{n-1}^\perp= \frac{\sqrt{\pi} \Gamma\left(\frac{n+3}{2(n+2)} \right)}{(n+1)\Gamma\left( \frac{1}{2(n+2)}\right)} \ ,
\ee
such that for $n=1$, $\lambda_1=1$, $c_0^\parallel=c_0^\perp=c_0$, $R=\ruv$. Then, the coefficients appearing in the EE are
\be\label{eq:AB}
\begin{split}
&A_\parallel=\frac{1}{\lambda_n^4} c_{n-1}^\parallel\ ,  \ B_\parallel= \frac{4}{n\lambda_n}c_{n-1}^\parallel \\
&A_\perp=\frac{1}{\lambda_n^4}c_{n-1}^\perp \ ,  \ B_\perp=\frac{2(n+1)}{\lambda_n} c_{n-1}^\perp\ .
\end{split}
\ee

In the IR limit $\ell\to\infty$, using the values of $d_\parallel=2/n$ and $d_\perp=n+1$ for the solutions with anisotropic scaling, the requirement that the $c$-functions asymptote to a constant value in the IR fixes
\be
 C_\parallel(\ell)\simeq \beta_{d_\parallel+2} \ell_0^3 \left(\frac{\ell}{\ell_0}\right)^{1+\frac{2}{n}}\ , \  C_\perp(\ell)\simeq \beta_{d_\perp+2} \ell_0^3 \left(\frac{\ell}{\ell_0}\right)^{n+2},\ \ell\to \infty\,, 
\ee
where $\ell_0$ is a scale fixed by the properties of the RG flow. We have chosen the coefficients according to the expected behavior for a conformal theory of dimensions $D=d+2$ \eqref{eq:cfuncD}. Then,
\bea
 &&\displaystyle{\lim_{\ell\to \infty}} c_\parallel =   \cuv\left( \frac{\mu R^2}{\ell_0}\right)^{\frac{2}{n}-2}\beta_{\frac{2}{n}+2}\frac{R^8}{\ruv^8}\frac{8}{n\pi}A_\parallel B_\parallel^{2/n} = \cuv\left(\frac{4}{n+3}\right)^2\frac{\lambda_n^{8-d_\parallel}\pi^{\frac{d_\parallel-2}{2}}}{n^{d_\parallel}\Gamma\left(\frac{1}{n} \right)}\left( \frac{\mu R^2}{\ell_0}\right)^{\frac{2}{n}-2}\\
 &&\displaystyle{\lim_{\ell\to \infty}} c_\perp = \cuv \left( \frac{\mu R^2}{\ell_0}\right)^{n-1} \beta_{n+3}\frac{R^8}{\ruv^8} \frac{4(n+1)}{\pi}A_\perp B_\perp^{n+1}  = \cuv\left(\frac{4}{n+3}\right)^2\frac{\lambda_n^{8-d_\perp}\pi^{\frac{d_\perp-2}{2}}}{\Gamma\left(\frac{n+3}{2} \right)} \left( \frac{\mu R^2}{\ell_0}\right)^{n-1} \ .
\eea
There is a combination that is independent of $\ell_0$. Let us define the averaged $c$-function
\be\label{eq:caveganiso}
\bar{c}=\left( c_\parallel^n c_\perp^2\right)^{\frac{1}{n+2}} \ .
\ee
Then,
\be
\lim_{\ell \to 0} \bar{c}=\cuv \ , \ \lim_{\ell \to \infty} \bar{c}= \cuv\left(\frac{4}{n+3}\right)^2\frac{\lambda_n^6}{\left(n^2 \Gamma\left(\frac{1}{n} \right)^n\Gamma\left(\frac{n+3}{2} \right)^2\right)^{\frac{1}{n+2}}} > \cuv\ .
\ee
If we consider $c_\parallel$ and $c_\perp$ separately, the most natural choice of scale seems to be $\ell_0=\mu R^2$ as other choices increase the value of either $c_\parallel$ or $c_\perp$. 
\begin{figure}[!ht]
\center
 \includegraphics[width=0.48\textwidth]{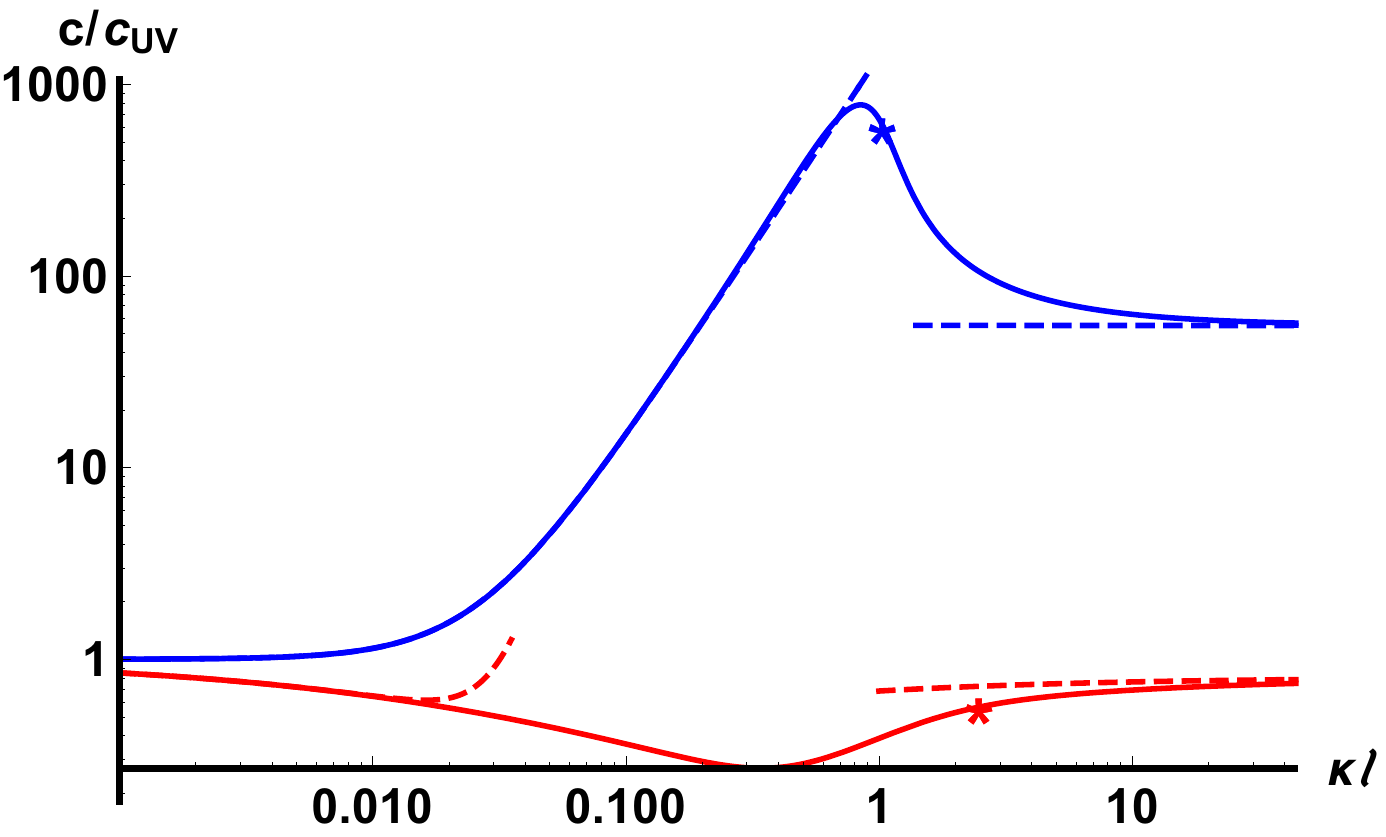}
 \includegraphics[width=0.48\textwidth]{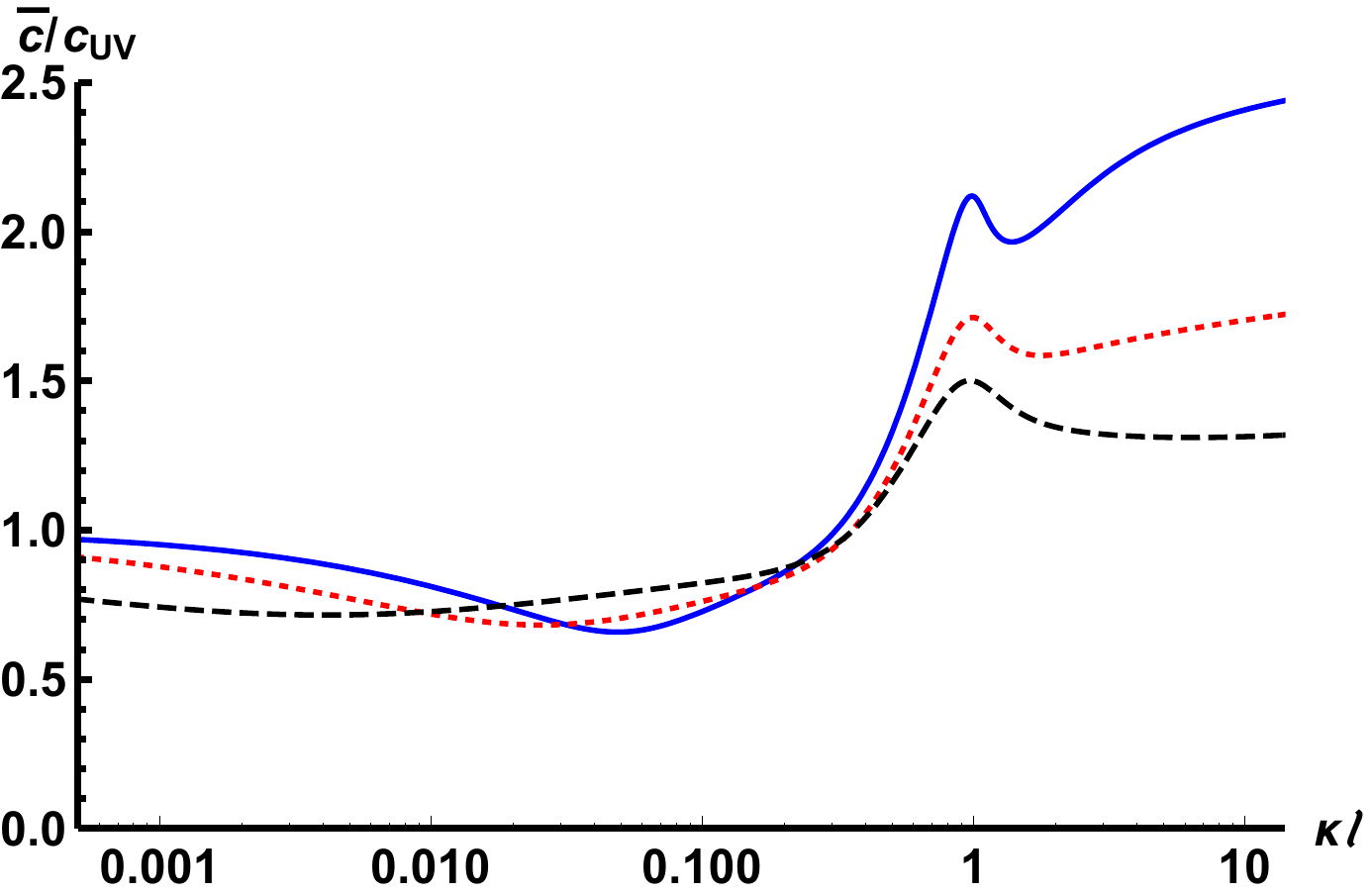}
  \caption{Left: The $c$-functions for the in-plane ($\parallel$,blue) and off-plane ($\perp$,red) directions at $Q_f=1$ and $n=1/2$, $m=2$ as functions of $\ell_\parallel$ and $\ell_\perp$, respectively. The dashed curves are the analytic UV and IR expansions.  Right: We depict the average $\bar c$-functions (\ref{eq:caveganiso}) for $n=1/3$ (solid blue), $n=1/2$ (dotted red), and $n=2/3$ (dashed black) for $Q_f=1$, $m=2$, and $\mu R^2/\ell_0=1$. Notice the log-linear scale.}
\label{fig:cEEaniso}
\end{figure}
With this choice, a direct evaluation gives, for any $1>n\geq 1/3$,
\be
 \lim_{\ell\to \infty} c_\parallel > \cuv> \lim_{\ell\to \infty} c_\perp \ .
\ee
We note that the hierarchy has switched with respect to the boomerang flows. It would be interesting to understand this phenomenon. Related to this, in the IR $c_\parallel>c_\perp$, and  $\cuv>c_\perp$ so $c_\perp$ is a candidate for a monotonically decreasing $c$-function. However, there is no unambiguous choice for the functions $C_\parallel(\ell)$ and $C_\perp(\ell)$, and the behavior of the $c$-functions at intermediate scales will depend on this choice. A simple possibility is
\be
C_\parallel(\ell)=\beta_4 \ell^3+\beta_{d_\parallel+2} \ell_0^3 \left(\frac{\ell}{\ell_0}\right)^{1+\frac{2}{n}}\ , \ C_\perp(\ell)=\frac{\beta_4 \ell^3}{1+\frac{\beta_4}{\beta_{d_\perp+2}} \left(\frac{\ell}{\ell_0} \right)^{1-n }} \ .
\ee
However, as in the boomerang case, we find that neither individual $c$-functions nor the averages (\ref{eq:caveganiso}) are monotonic, see Fig.~\ref{fig:cEEaniso}. Instead, they peak roughly at the intrinsic energy scale of the background. 


\subsection{$c$-function from null congruences}

In this section we present an alternative holographic $c$-function for our models, following the proposal of \cite{Sahakian:1999bd}, based on ideas of \cite{Bousso:1999xy} , which proposed to use the expansion parameter of the congruences of null geodesics to extract the information encoded holographically in the geometry (see also \cite{Alvarez:1998wr} for a similar proposal for the $c$-function).  For a 4d QFT the $c$-function of \cite{Sahakian:1999bd} is defined by the geodesics of its 5d dual geometry. The corresponding metric for our case can be obtained by reducing the Ansatz (\ref{metric_ansatz_zeta}) to five dimensions. This metric reads as follows
\be\label{reduced_5d_metric}
 ds^2_5 = \zeta^{{8\over 3}}\,h^{{1\over 3}}\,e^{{2\over 3}\,f}\,\Big[-(dx^0)^2+(dx^1)^2+(dx^2)^2+e^{-2\phi}\,(dx^3)^2\Big]+\zeta^{{14\over 3}}\,h^{{4\over 3}}\,e^{-{4\over 3}\,f}\,d\zeta^2\ .
\ee
The first step in the proposal of \cite{Sahakian:1999bd} is to consider a null vector $k^{\mu}$ tangent to the geodesics of the type:
\be
 k^{\mu}=F(\zeta)\Bigg({1\over \sqrt{| g_{x_0 x_0 }|}}\partial_{x_0}-{1\over \sqrt{ g_{\zeta \zeta }}}\partial_{\zeta}\Bigg)=F(\zeta)\Big(\zeta^{-{4\over 3}}\,h^{-{1\over 6}}\,e^{-{1\over 3}\,f} \partial_{x_0}\,-\,\zeta^{-{7\over 3}}\,h^{-{2\over 3}}\,e^{{2\over 3}\,f}\partial_{\zeta}\Big)\ ,
\ee
where the function $F(\zeta)$ is obtained by imposing the affine condition:
\be
 k^{\mu}\,\nabla_{\mu}\,k^{\nu} = 0\ .
\ee
It is easy to see that, in our geometry (\ref{reduced_5d_metric}),  the function $F(\zeta)$  must satisfy the following differential equation:
\be
 {F'\over F}\,=\,-{4\over 3\,\zeta}\,-\,{1\over 6}\,{h'\over h}\,-\,{1\over 3}\,f'\ ,
\ee
which can be integrated as
\be
 F(\zeta)\,=\,\zeta^{-{4\over 3}}\,h^{-{1\over 6}}\,e^{-{1\over 3}\,f}\ .
\ee
Thus, the vector $k^{\mu}$ becomes:
\be
 k^{\mu}= \zeta^{-{8\over 3}}\,h^{-{1\over 3}}\,e^{-{2\over 3}\,f} \partial_{x_0}\,-\,\zeta^{-{11\over 3}}\,h^{-{5\over 6}}\,e^{{1\over 3}\,f}\partial_{\zeta}\ .
\ee
The expansion parameter $\theta$ for the congruence is defined as
\be
 \theta = \nabla_{\mu}\,k^{\mu}\ .
\ee
This parameter measures the isotropic expansion of the flow of null geodesics  in the geometry. In our metric $\theta$ takes the form:
\be
 \theta = -{1\over 2}\,\zeta^{-{14\over 3}}\, h^{-{11\over 6}}\,e^{{1\over 3}\,f}\Big(\zeta\,h'\,+\,2\,h(4+\zeta f'\,-\,\zeta \phi')\Big)\ .
\ee
In the proposal of \cite{Sahakian:1999bd} the holographic central charge is given by:
\be\label{c_congruence_proposal}
 c(\zeta)\sim {1\over \sqrt{H}\,\theta^{3}}\ ,
\ee
where $H$ is the determinant of the induced metric on hypersurfaces with constant $x^0$ and $\zeta$. In our case it is straightforward to check from (\ref{reduced_5d_metric}) that
$\sqrt{H} = \zeta^{4}\,h^{{1\over 2}}\,e^{f-\phi}$.
Therefore we can write $c(\zeta)$ as:
\bea
c(\zeta) & = & {432\over Q_c^2}{\zeta^{10}h^{5}e^{\phi-2f}\over \Big(\zeta h'+2h(4+\zeta f'-\zeta \phi')\Big)^3}\cuv =  {3456\over Q_c^2}{\zeta^{31}h^{5}e^{4f-5\phi}\over \Big[\Big(\zeta^{8}he^{2f-2\phi}\Big)'\Big]^3}\cuv \label{congruence_c_ani_compact} \ ,
\eea
where  we have absorbed the multiplicative constant of (\ref{c_congruence_proposal}) in  $\cuv=c(\zeta\to\infty)$.  
\begin{figure}[ht]
\center
 \includegraphics[width=0.48\textwidth]{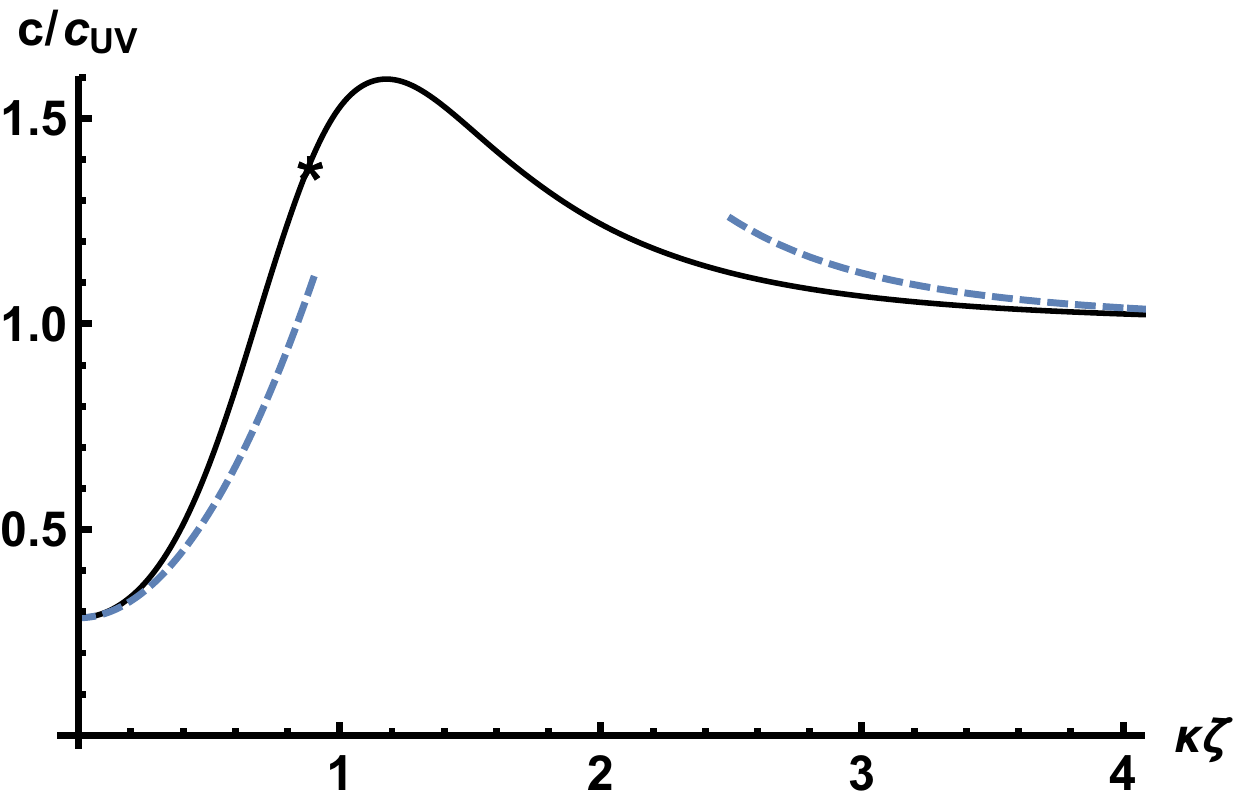}
 \includegraphics[width=0.48\textwidth]{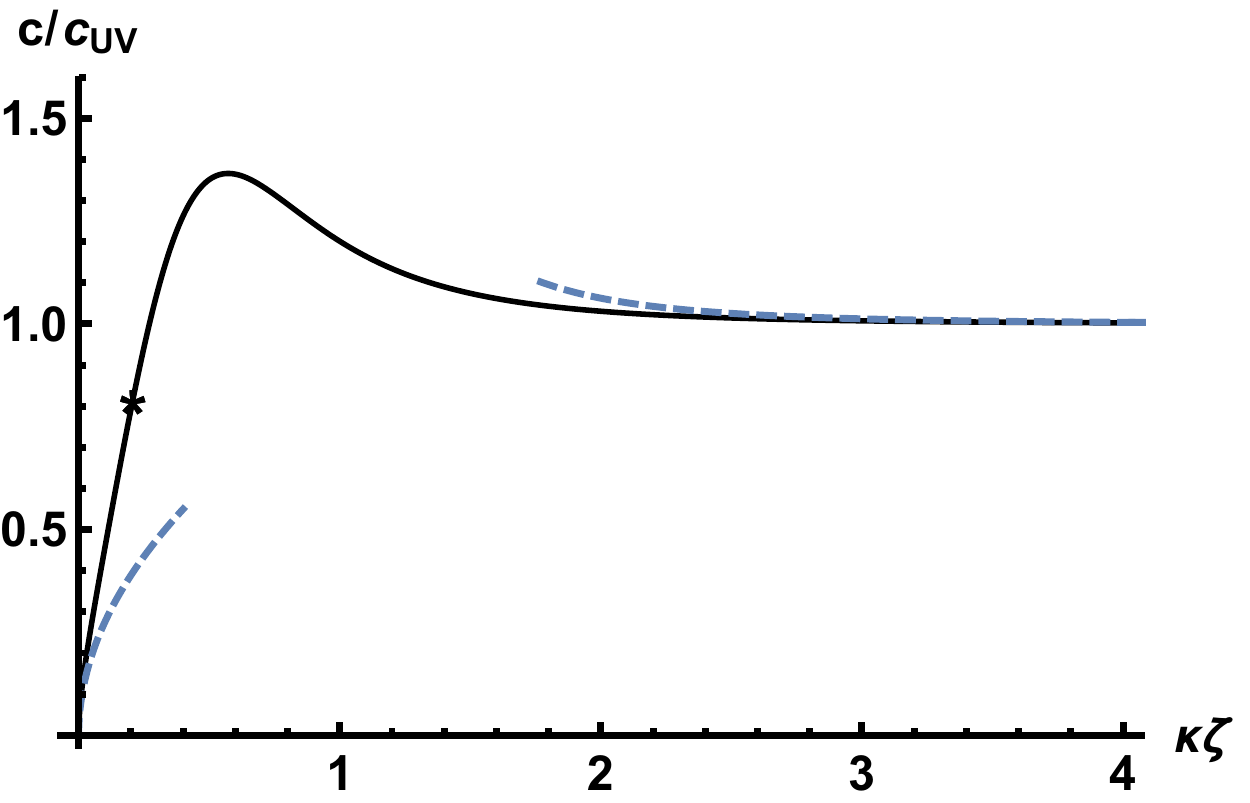}
  \caption{Left: We depict $c(\zeta)/\cuv$ for a boomerang solution with $n=3$ and $m=2$ with $Q_f=10$. Right: We depict the same quantity for an anisotropic IR solution with $n=1/2$ and $m=2$ with $Q_f=1$. The dashed curves correspond to the analytic UV and IR asymptotics (\ref{c_UV}) and (\ref{eq:c_IR}), respectively.}\label{fig:congruence_c}
\end{figure}

We have plotted in Fig.~\ref{fig:congruence_c} the function $c(\zeta)$ for boomerang and anisotropic Lifshitz flows (\ref{congruence_c_ani_compact}). We find qualitatively similar results to the ones found by using entropic $c$-functions. In particular, notice that in all cases, $c(\zeta)$ is never monotonic in the whole range of $\zeta$. 

Furthermore, we find that the UV behavior of $c(\zeta)$ is universal and given by, expanding out (\ref{congruence_c_ani_compact}),
\be\label{c_UV}
 c(\zeta)  =  \cuv\Big[1+ {Q_f\over (\kappa\,\zeta)^4} +\ldots\Big] \ , \  \zeta\to\infty \ .
\ee
Notice, in particular, that this means that $c(\zeta)$ {\emph{decreases}} as the UV is approached. As usual, the IR behavior is different for boomerang and anisotropic Lifshitz flows. 
We find, as $\zeta\to 0$,
\be\label{eq:c_IR}
 c(\zeta) \simeq \left\{ \begin{array}{ll}
   {\cuv\over w_{n,m}}\Big[1+{n^2+10n+1\over 2(5+n)(n-1)}{Q_f\over  w_{n,m}}(\kappa\zeta)^{n-1}\Big]   & , \ \ n>1 \\
   2\,\cuv{(1-n)(5+n)^3\over (2+n)^3(3+n)^2Q_f}(\kappa\zeta)^{1-n}  & , \ \ n<1 \ .\end{array}\right. 
\ee
Recalling that $w_{n,m}>0$ we find that in all the cases the IR value is smaller than $\cuv$. In the anisotropic Lifshitz case, the IR value actually tends to zero.

\section{Discussion and outlook}\label{sec:discussion}

We constructed a new family of anisotropic solutions of ten-dimensional supergravity coupled to smeared brane sources. The solutions are supersymmetric and we argued that they are dual to ${\cal N}=4$ SYM with an expectation value for a three-form operator. This operator can be Hodge dualized to an axial current with a non-zero expectation value along the spatial direction transverse to the smeared branes. We did not determine whether the smeared sources we introduced can actually be realized microscopically within string theory, so our construction is not fully top-down. To go beyond supergravity and to address this point is an important extension of our work that deserves a more detailed study in the future.

From a more phenomenological point of view, an interesting aspect of the model is that the distribution of smeared branes is an almost arbitrary function of the holographic radial coordinate. In the cases we have studied, the desired IR behavior can be engineered by changing the density of branes in the horizon region. In this work, our family of solutions consist of boomerang flows between conformal fixed points and those that will have an anisotropic scaling in a spatial direction deep in the IR. 
In principle, it is possible to design a brane distribution in such a way that an intermediate scaling region appears, emulating other results in boomerang flows of \cite{Donos:2017ljs,Donos:2017sba}. Following this line of thought, more exotic possibilities such as flows where the density has several maxima in the radial direction are also open to study. Since the full geometry is  determined by simple formulas stockpiling the brane distribution, it becomes a straightforward exercise to construct new anisotropic solutions. All these solutions are supersymmetric and so the stability is guaranteed. This solution-generating technique leading to explicit and even analytic geometries is not commonplace in supergravity constructions.

Our analysis of the entanglement entropy and holographic $c$-functions shows that one should be careful when discussing monotonicity results for these quantities in holographic RG flows obtained in dimensionally reduced supergravities. If one identifies a functional for the entanglement entropy as the area of a codimension two surface, depending on the warp factors along the field theory directions  in domain wall coordinates, the corresponding functional in the reduced theory will generally be different than the area functional for the ten-dimensional metric when the internal space has non-trivial factors. Thus, we find that all existing proofs of monotonicity in the anisotropic case are not directly applicable to the ten-dimensional construction, and none of the usual proposals yield monotonic $c$-functions. We observe that the non-monotonic behavior is correlated with the profile of the brane distribution, which also determines the degree of anisotropy. In a certain sense the $c$-functions are sensitive to the number of degrees of freedom in the bulk, although this does not have a direct translation to the degrees of freedom in the dual field theory. It should be noted that similar non-monotonic behavior was observed in the boomerang supergravity solutions \cite{Donos:2017ljs,Donos:2017sba} and it is interesting to ask if a similar interpretation would apply in those cases, for instance in terms of background fluxes. In order to better understand the properties of the solutions along the full ten-dimensional anisotropic RG flow it would be interesting to study other observables that are also sensitive to the internal energy scales \cite{Bea:2013jxa,Balasubramanian:2018qqx,Jokela:2019ebz}, such as mutual information, entanglement wedge cross sections, or Wilson loops. As we have mentioned, for a large enough brane density, preliminary results indicate that some of these quantities could go through different saddle points as their size is varied.

Regarding other extensions, it would be very interesting to construct anisotropic black hole solutions, perhaps also including charge. Those would be dual to anisotropic states at finite temperature and charge density, and could be used as toy models of real anisotropic systems as alluded to in the introduction. Since supersymmetry will be broken, it is to be expected that stable configurations do not admit an arbitrary distribution of smeared branes, but rather that it will be unique or very constrained, if it exists. In this work we have focused on duals to states with spontaneously broken isotropy, but our identification of the dual operator sourced by the branes as an axial current connects the multilayered solutions of  \cite{Penin:2017lqt,Jokela:2019tsb,Gran:2019djz} to the physics of Weyl semimetals (see, {\emph{e.g.}}, \cite{Grushin:2012mt}), although in the last case the axial current is Abelian. It is clearly interesting to pursue this direction further.

Concerning other smeared brane configurations, we note here that in most cases the brane distribution can also be chosen almost arbitrarily, but so far this has not been explored much. This is partly because it is not easy (or maybe possible) to find localized brane configurations corresponding to a given distribution, so the construction becomes more phenomenological. Nevertheless, it would be interesting to explore other brane constructions that are Lorentz invariant, such as the D3-D7 intersection \cite{Benini:2006hh,Benini:2007gx,Nunez:2010sf}, in order to disentangle the effects of the anisotropy from other properties of the smeared brane construction.


\addcontentsline{toc}{section}{Acknowledgments}
\paragraph{Acknowledgments}
We would like to thank Valentina Giangreco M. Puletti, Blaise Gout\'eraux, Oscar Henriksson, Elias Kiritsis, Carlos Nunez, Juan F. Pedraza, Giuseppe Policastro, Mehri Rahimi, Christopher Rosen, and Javier Tarr\'io for discussions. C.~H. is partially supported by the Spanish grant PGC2018-096894-B-100 and by the Principado de Asturias through the grant GRUPIN-IDI/2018 /000174. N.~J. is supported in part by the Academy of Finland grant no. 1322307.  J.~M.~P. and A.~V.~R. are funded by the Spanish grant FPA2017-84436-P, by Xunta de Galicia-Conseller\'\i a de Educac\' \i on  (ED431C-2017/07), by FEDER and by the Maria de Maeztu Unit of Excellence MDM-2016-0692.  J.~M.~P. is also supported by the Spanish FPU fellowship FPU14/06300 and a Royal Society University Research Fellowship Enhancement Award.

\appendix

\section{Background details}\label{Background_details}

In this appendix we flesh out more details of the family of backgrounds found in \cite{Conde:2016hbg,Penin:2017lqt,Jokela:2019tsb} and generalize in the current context. 
Besides the metric and the dilaton written in (\ref{metric_ansatz_zeta}) and (\ref{f_phi_W}), these backgrounds of type IIB supergravity contain a 
RR five-form $F_5$ and a RR three-form $F_3$. The former is self-dual and given by the standard Ansatz in terms of the dilaton $\phi$ and warp factor $h$:
\be\label{F5_sol}
 F_5 = \partial_{\zeta}\,\big(e^{-\phi}\,h^{-1}\big)\,\big(1+*\big)\,d^4x\wedge d\zeta\ .
\ee
In order to write the expression for $F_3$, let us recall that the  ${\mathbb C}{\mathbb P}^2$  manifold is a K\"ahler-Einstein space endowed with a K\"ahler two-form $J=dA/2$,  where the one-form potential $A$ is the one appearing in the $U(1)$ fibration of the metric (\ref{metric_ansatz_zeta}).  The two-form $J$ can be canonically written as $J=e^1\wedge e^2+e^3\wedge e^4$, where $e^1,\ldots,e^4$ are vielbein one-forms of ${\mathbb C}{\mathbb P}^2$, whose explicit coordinate expressions can be found in appendix A of \cite{Conde:2016hbg}. Let us introduce the complex two-form $\hat\Omega_2$ as
\be\label{hat_Omega_2}
 \hat\Omega_2\,=\,e^{3 i\tau}\,(e^1+i e^2)\wedge (e^3+i e^4) \ .
\ee
Then,  we can write $F_3$ as follows
\be\label{F3_ansatz}
 F_3\,=\,Q_f\,p(\zeta)\,dx^3\wedge {\rm Im }\,\hat\Omega_2\ ,
\ee
where $Q_f$ is a constant and $p(\zeta)$ is an arbitrary function of the holographic coordinate $\zeta$. Clearly, $d\,F_5=0$, since the D3-branes have been replaced by a flux in the supergravity solution. However, $dF_3\not=0$,  which means that the Bianchi identity for $F_3$ is violated due to the presence of the D5-branes. By inspecting the expression of $dF_3$ we immediately conclude that we are continuously distributing D5-branes along the $x^3$ direction, giving rise to a system of multiple $(2+1)$-dimensional parallel layers. This is, of course, the origin of the anisotropy of the backreacted metric. The function $p(\zeta)$ determines the D5-brane charge distribution in the holographic direction. This background is supersymmetric and satisfies the equations of motion of supergravity with delocalized D5-brane sources if $W$ satisfies (\ref{Master_W}) and $\phi$, $f$, and $h$ are given in terms of $W$ as in (\ref{f_phi_W}) and (\ref{h_W}). 

Let us derive the expression for $h$ written in (\ref{h_W}). It was shown in \cite{ Conde:2016hbg,Jokela:2019tsb} that the warp factor is the solution of the following first-order differential equation
\be\label{warp_factor_eq}
 {dh\over d\zeta}\,+\,Q_f\,{e^{{3\phi\over 2}-f}\,p\over \zeta}\,h\,=\,-{Q_c\over \zeta^3}\,e^{-2f}\ .
\ee
Let us proceed solving (\ref{warp_factor_eq}) in general, in terms of an arbitrary $W$.  We first use (\ref{f_phi_W}) to  write the coefficient multiplying $h$ in (\ref{warp_factor_eq}) in terms of $W$
\be\label{second_term_h_equation}
 Q_f\,{e^{{3\phi\over 2}-f}\,p\over \zeta}\,=\,{1\over W\,+\,{1\over 6}\,\zeta\,{d W\over d\zeta}}\,{Q_f\,p(\zeta)\over \zeta^2\,\sqrt{W}}\ .
\ee
Using the master equation (\ref{Master_W}), the right-hand side of (\ref{second_term_h_equation}) can be written as a total derivative
\be
 Q_f\,{e^{{3\phi\over 2}-f}\,p\over \zeta}\,=\,-{d\over d\zeta}\log\Big[\,W\,+\,{1\over 6}\,\zeta\,{d W\over d\zeta} \Big]\ .
\ee
Moreover, since
\be
 {e^{-2f}\over \zeta^3}\,=\,{1\over \zeta^{5}\,W}\,\Big[\,W\,+\,{1\over 6}\,\zeta\,{d W\over d\zeta} \Big]\ ,
\ee
the equation determining $h$ is:
\be
{dh\over d\zeta}\,-\,{d\over d\zeta}\log\Big[W\,+\,{1\over 6}\,\zeta\,{d W\over d\zeta} \Big]\,h\,=\,-\,{Q_c\over \zeta^{5} \,W}\,\Big[\,W\,+\,{1\over 6}\,\zeta\,{d W\over d\zeta} \Big]\ .
\ee
We can solve this differential equation by variation of constants. To start with, notice that formally when $Q_c\to 0$, the differential equation becomes homogeneous and the solution is readily obtained
\be
 h(\zeta) = C\,\Big[\,W\,+\,{1\over 6}\,\zeta\,{d W\over d\zeta} \Big]\ , \  Q_c=0 \ ,
\ee
where $C$ is a constant. Next, we allow $C$ to depend on $\zeta$ and substitute it into the original differential equation, yielding a differential equation for $C(\zeta)$:
\be
 {d\, C\over d\zeta}\,=\,-{Q_c\over \zeta^5\,W(\zeta)}\ .
\ee
This is simply integrated to
\be
 C(\zeta)\,=\,Q_c\,\int_{\zeta}^{\zeta_0}{d\bar \zeta\over \bar\zeta^5\,W(\bar\zeta)}\ ,
\ee
where $\zeta_0$ is a constant of integration. Finally, let us choose $\zeta_0$ in such a way that $h(\zeta\to\infty)=0$. This then brings us to
\be
 h(\zeta) = Q_c \Big[W(\zeta)\,+\,{\zeta\over 6}\,{d W\over d\zeta} \Big]\, \int_{\zeta}^{\infty}{d\bar \zeta\over \bar\zeta^5\,W(\bar\zeta)}\ .
\ee
Taking into account the expression of the dilaton in (\ref{f_phi_W}), we land on (\ref{h_W}).

\subsection{Solution to the master equation}

Let us now show how we integrate the master equation (\ref{Master_W}) in general. First of all, we define a new function $F(\zeta)$ as follows
\be\label{F_def}
 F(\zeta) \equiv {p(\zeta)\over \sqrt{W(\zeta)}}\ .
\ee
Then, it is straightforward to demonstrate that the master equation becomes
\be\label{master_eq_with_F}
 {d\over d\zeta}\Big(\zeta^7\,{d W\over d\zeta}\Big)\,=\,-6\,Q_f\,\zeta^4\,F(\zeta)\ .
\ee
Given the structure of the left-hand side of (\ref{master_eq_with_F}), we can simply perform a double integration
\be\label{W_sol_double_int}
 W(\zeta)\,=\,1\,+\,6\,Q_f\,\kappa\,\int_{\kappa\,\zeta}^{\infty}\,{dx\over x^7}\,\int_0^{x}\,u^4\,F\Big({u\over \kappa}\Big)\,du\ .
\ee
In (\ref{W_sol_double_int})  $\kappa$ is an arbitrary constant and we have already imposed that $W(\zeta\to\infty)=1$. Integrating by parts in the integral over $x$ in (\ref{W_sol_double_int}), and assuming that $x^{-1}\,F(x)\to 0$ as $x\to\infty$, we can rewrite (\ref{W_sol_double_int}) as a single integral
\be\label{W_sol_single_int}
 W(\zeta)\,=\,1+{Q_f\kappa\over (\kappa \zeta)^6}\,\int_0^{\kappa\zeta}\,dx\,x^4\,F(x/\kappa)\,+\,Q_f\kappa\,\int_{\kappa\zeta}^{\infty}\,dx\,{F(x/\kappa)\over x^2}\ .
\ee
As a check one can directly show that (\ref{W_sol_single_int}) solves (\ref{master_eq_with_F}). 

The profile function (\ref{general_profile}) we use to generate our geometries corresponds to the following explicit expression for $F$:
\be\label{F_n_m}
 F(x/\kappa)\,=\,{1\over \kappa}\,{x^n\over (1+x^m)^{{n+3\over m}}}\ .
\ee
Plugging (\ref{F_n_m}) into (\ref{W_sol_single_int}) we arrive at the following integrals
\be\label{W_n_m_integrals}
W(\zeta)\,=\,1\,+\,{Q_f\over   (\kappa\zeta)^6}\,\int_0^{\kappa\zeta}\,dx\, {x^{n+4}\over (1+x^m)^{{n+3\over m}}}\,+\,Q_f\,\int_{\kappa\zeta}^{\infty}\,dx\, {x^{n-2}\over (1+x^m)^{{n+3\over m}}}\ .
\ee
The integrals in (\ref{W_n_m_integrals}) can be done analytically in terms of hypergeometric functions, giving (\ref{W_general_sol}). Finally, for expansions at the IR, it is useful to rewrite $W$ as
\bea
&&W(\zeta)=1+{\Gamma\Big({4\over m}\Big)\,\Gamma\Big({n-1\over m}\Big)\over m\,
\Gamma\Big({3+n\over m}\Big)}\,Q_f\,+\, Q_f\,(\kappa\zeta)^{n-1}\,\Bigg[{1\over n+5}
F\Big({5+n\over m}, {3+n\over m}; {5+m+n\over m};- (\kappa \zeta)^m\Big) \rc\rc
&&\qquad\qquad\qquad\qquad\qquad\qquad
+{1\over 1-n}\,F\Big({n-1\over m}, {3+n\over m}; {m+n-1\over m};- (\kappa \zeta)^m\Big)\Bigg]\,\,,
\eea
while for expansions near the boundary we instead use
\bea
&&W(\zeta)=1+{1\over 2}\,{Q_f\over (\kappa\zeta)^4}\,
\Bigg[F\Big(-{2\over m}, {3+n\over m}; {m-2\over m};- (\kappa \zeta)^{-m}\Big)
\qquad\qquad\qquad\qquad\qquad\qquad
\rc\rc
&&\qquad\qquad\qquad
+{1\over 2}F\Big({4\over m}, {3+n\over m}; {4+m\over m};- (\kappa \zeta)^{-m}\Big)\Bigg]
+{\Gamma(-{2\over m})\,\Gamma\Big({5+n\over m}\Big)\over m\,
\Gamma\Big({3+n\over m}\Big)}\,{Q_f\over (\kappa\zeta)^6}\ .
\eea

\subsection{Reduction to five dimensions}

Let us lay out the dimensional reduction of our system to a gravity theory in five dimensions. We will not write down all the details explicitly, but will refer to key formulas in the literature. The reduction Ansatz for the metric has been written in (\ref{10d_5d_metric_ansatz}).  In the reduced 5d theory we have three scalars,$\gamma$ and $\lambda$ for the metric  (\ref{10d_5d_metric_ansatz}) and the dilaton $\phi$. In order to match the metric (\ref{10d_5d_metric_ansatz}) with the Ansatz (\ref{metric_ansatz_zeta}) we need to relate $h$, $f$, and $\zeta$ to ($\gamma$,$\lambda$) and to one of the components of the 5d metric $g_{pq}$.  For convenience we choose the $g_{\zeta\zeta}$ component as the independent function.  It can be easily verified that the seeked relation is
\be
 h^{{1\over 2}}\,=\,e^{{10\over 3}\,\gamma+10\lambda}\,g_{\zeta\zeta}\ , \ e^{f}\,=\,{e^{-{8\gamma\over 3}-\lambda}\over \sqrt{g_{\zeta\zeta}}} \ , \ \zeta\,=\,{e^{-{8\gamma\over 3}-6\lambda}\over \sqrt{g_{\zeta\zeta}}}\ ,
\ee
which can be inverted as:
\be\label{eq:5dscalars}
 e^{\lambda} =  \zeta^{-{1\over 5}}\,e^{{f\over 5}}\ , \ e^{\gamma} = \zeta^{-{4\over 5}}\,h^{-{1\over 4}}\,e^{-{f\over 5}} \ , \ g_{\zeta\zeta}\,=\,h^{{4\over 3}}\,\zeta^{{14\over 3}}\,e^{-{4f\over 3}} \ .
\ee
The reduced 5d theory also contains a four-form ${\cal F}_4$ which originates from the reduction of the RR three-form $F_3$ of ten-dimensional supergravity. Moreover, our system also contains dynamical D5-branes, which are codimension one objects in the reduced 5d theory, extended along the hypersurface $x^3={\rm constant}$ and then smeared over $x^3$. The corresponding DBI action contains the determinant of the induced metric on this 4d surface, which we will denote by $\hat g_4$, integrated over $x_3$ to account for the smearing. The full effective action can be obtained by generalizing the results in \cite{Penin:2017lqt}, yielding
\bea
&&S_{eff}\,=\,{V_5\over 2\,\kappa_{10}^2}\,
\int d^5z\, \sqrt{-g_5}\,\,\Big[
R_5\,-\,{40\over 3}\,(\partial\gamma)^2\,-\,20\,(\partial\lambda)^2\,-\,
{1\over 2}\,(\partial\phi)^2\,-\,
{1\over 2\cdot 4!}\,e^{-4\gamma-4\lambda-\phi}\,( {\cal F}_4)^2\,\rc\rc
&&\qquad\qquad\qquad\qquad\qquad\qquad\qquad\qquad
-U_{scalars}\,\Big]\,+\,S_{branes}\,+\,S_{WZ}
\,\,,
\label{effective_5d_action}
\eea
where $V_5$ is the volume of the five dimensional compact space and
$U_{scalars}$ is the  following potential for $\lambda$ and $\gamma$:
\be
U_{scalars}=4\,e^{{16\over 3}\,\gamma+12 \lambda}-24\,e^{{16\over 3}\gamma+2\lambda} +{Q_c^2\over 2}\,e^{{40\over 3}\gamma} \,\,.
\label{5d_potential}
\ee
The construction of the action $S_{WZ}$ will be addressed later, starting at around (\ref{eq:5dWZ}). In order to find $S_{branes}$ we proceed as in appendix C of \cite{Penin:2017lqt} and look at the DBI action of the distribution of D5-branes. For a calibrated set of smeared branes the resulting DBI action equals (minus) the WZ one which is the integral of the wedge product of the RR potentials and the smearing form $\Xi$. In our case the relevant RR potential is the six-form $C_{6}$ and so the corresponding action is
\be\label{Branes_smeared}
 S_{branes} = -T_5 \int_{{\cal M}_{10}} C_{6}  \wedge \Xi\ ,
\ee
where $\Xi$ is a four-form. The expressions for $C_{6}$ and $\Xi$  are given in appendix B.2 of \cite{Jokela:2019tsb}. After integrating over the angular directions, we can rewrite (\ref{Branes_smeared}) as:
\be
 S_{branes}\,=\,\int d x^0 \,dx^1\,dx^2\,dx^3\,d\zeta\,{\cal L}_{branes}\ ,
\ee
where ${\cal L}_{branes}$ is a smeared  Lagrangian density. Using the results in \cite{Jokela:2019tsb}, ${\cal L}_{branes}$  reads
\be\label{smeared_massive}
 {\cal L}_{branes}\,=\,-{V_5\,Q_f\,\over \kappa_{10}^2}\,\zeta^3\,e^{{\phi\over 2}-f}\,\Big(3\,p(r)\,+\,{e^{2f}\over \zeta}\,{dp\over d\zeta}\Big)\ .
\ee
Let us now rewrite this last expression in a covariant form with respect to the $5d$ metric $g_{pq}$. First of all, we notice that the function multiplying $dp/d\zeta$ in (\ref{smeared_massive}) can be written as:
\be
 {e^{2f}\over \zeta}\,=\,{e^{4\lambda-{8\gamma\over 3}}\over \sqrt{g_{\zeta\zeta}}}\ .
\ee
Second, the determinant $\hat g_4$ of the induced metric in the $x^3=0$ submanifold spanned by the D5-branes is related to $\gamma$, $\lambda$,  and $g_{\zeta\zeta}$ as
\be
 \sqrt{-\hat g_4}\,=\,{e^{-10\gamma-15\lambda}\over g_{\zeta\zeta}}\ .
\ee
As a consequence, we can rewrite the prefactor in (\ref{smeared_massive}) as:
\be
 \zeta^3\,e^{-f}\,=\,e^{{14\gamma\over 3}\,-\,2\lambda}\,\sqrt{-\hat g_4}\ .
\ee
Putting all these results together, we can write the brane action in (\ref{effective_5d_action}) as:
\be
 S_{branes}\,=\,-{V_5\over 2\,\kappa_{10}^2}\,\int d^5z\,\sqrt{-\hat g_4}\,U_{branes}\ ,
\ee
where $U_{branes}$ is the following function depending on the profile $p$:
\be\label{U_branes_5d}
 U_{branes}\,=\,2\,Q_f\,e^{{\phi\over 2}\,-\,2\lambda\,+{14\gamma\over 3}}\,\Big(3p+{e^{4\lambda-{8\gamma\over 3}}\over  \,\sqrt{g_{\zeta\zeta}}}\,{dp\over d\zeta}\Big)\ .
\ee

In order to write $U_{branes}$ in a covariant form, let us next introduce a vector field $v^{n}$ with unit norm in the $5d$ metric
\be
 v_p\,v^{p}\,=\,g_{pq}\,v^p\,v^{q}\,=\,1\ .
\ee
When $v^p$ points in the radial direction, only $v^{\zeta}$ is non-vanishing and given by
\be\label{v_vector_radial}
 v^p\,=\,{1\over \sqrt{g_{\zeta\zeta}}}\,\delta_\zeta^p\ .
\ee
In this case, we have
\be
 {\partial_\zeta\,p\over \sqrt{g_{\zeta\zeta}}}\,=\,v^n\,\partial_n\,p\,\equiv \nabla_v\,p\ ,
\ee
where $\nabla_v$ is the directional derivative along the unit vector $v$. It follows that $U_{branes}$ can be written as
\be\label{U_branes_5dv2}
 U_{branes} = 6\,Q_f\,e^{{\phi\over 2}\,-\,2\lambda\,+{14\gamma\over 3}}\,\Big(p+{e^{4\lambda-{8\gamma\over 3}}\over 3}\,\nabla_v\,p\Big)\ .
\ee

Let us finally discuss the ingredients in describing $S_{WZ}$. Let us define the one-form ${\cal F}_1$ via 5d Hodge dual of ${\cal F}_4$ as
\be\label{def_F4}
 {\cal F}_1\,=\,-e^{-4\gamma-4\lambda-\phi}\,*\,{\cal F}_4\ .
\ee
The one-form ${\cal F}_1$ is the result of reducing the RR 10d three-form $F_3$ to 5d, which is not closed and thus violates the Bianchi identity due to the presence of D5-brane sources. We thus expect to have $d{\cal F}_1\not=0$ in the reduced theory. As in the 10d formalism, the violation of Bianchi identity is induced by a Wess-Zumino term in the action (\ref{effective_5d_action}). It is easy to conclude that this term must have the form
\be\label{eq:5dWZ}
 S_{WZ}\,=\,{V_5\over 2\,\kappa_{10}^2}\,\int {\cal C}_3\wedge \Sigma_2\ ,
\ee
where ${\cal C}_3$ is the three-form potential for ${\cal F}_4$ and $\Sigma_2$ is a smearing two-form. Indeed, from the equation of motion for ${\cal C}_3$ (\ref{eom_F4_form}) one readily gets
\be
 d\,{\cal F}_1\,=\,\Sigma_2\ ,
\ee
which is the desired modified Bianchi identity. For our BPS Ansatz we have:
\be\label{F_1_ansatz}
 {\cal F}_1\,=\,\sqrt{2}\,Q_f\,p(\zeta)\,dx^3\ ,
\ee
and the smearing two-form $\Sigma_2$ is the one written in (\ref{Sigma_2}).

Next, let us look at the equations of motion that follow from the action (\ref{effective_5d_action}). The equation for the three-form ${\cal C}_3$  has been studied in Sec.~\ref{sec:fieldtheory}, cf. (\ref{eom_F4_form}). In order to write compactly the equations for the scalars, let us group them in a three-component field 
$\Psi=(\phi, \gamma, \lambda)$. Then, if $\alpha_\phi$, $\alpha_\gamma$, and $\alpha_\lambda$ take the values 
\be\label{alpha_Psi}
 (\alpha_{\phi}\,,\,\alpha_{\gamma}\,,\,\alpha_{\lambda}) = \left(1\,,\,{3\over 80}\,,\,{1\over 40}\right) \ ,
\ee
then the equations of motion of the scalars are
\be\label{5d_scalar_eoms_compact}
 \Box \Psi\,=\,\alpha_{\Psi}\,\partial_{\Psi}\,U_{scalars}\,+\,{1\over 2}\,\alpha_{\Psi}\,\big( {\cal F}_1\big)^2\,\partial_{\Psi}\, \Big(e^{4\lambda+4\gamma+\phi}\Big)
+ {\sqrt{-\hat g_4}\over \sqrt{-g_5}}\,\alpha_{\Psi}\,\partial_{\Psi} U_{branes} \ .
\ee
The Einstein equations are obtained by computing the variation of the action with respect to the 5d metric. The result is
\bea
R_{pq}-{1\over 2}\,g_{pq}\,R & = & \sum_{\Psi}\,{1\over 2\alpha_{\Psi}}\,\Big(\partial_p\,\Psi\,\partial_q\,\Psi\,-\,{1\over 2}\,g_{pq}\,(\partial \Psi)^2\Big)\,-\,{1\over 2}\,g_{pq}\,U_{scalars} \label{Einstein_eqs_5d} \\
& & +{1\over 2\cdot 4!}\, e^{-4\gamma-4\lambda-\phi}\Big(4\big({\cal F}_4\big)_{p\,r_1\, r_2\, r_3} \big({\cal F}_4\big)_q^{\,\,\,r_1\, r_2\, r_3}-\,{1\over 2}\,g_{pq}\,\big( {\cal F}_4\big)^2\Big) +T_{pq}^{branes} \nonumber \ ,
\eea
where $T_{pq}^{branes}$ represents the contribution originating from the brane term (\ref{U_branes_5dv2}). The non-vanishing components of $T_{pq}^{branes}$ are
\bea
T_{x^\mu x^\nu}^{branes} & = & -{Q_f\,e^{{3\phi\over 2}-7\lambda+{4\gamma\over 3}}\over \sqrt{g_{\zeta\zeta}}}\,\Big(3\,p\,+\,{e^{4\lambda-{8\gamma\over 3}}\over \sqrt{g_{\zeta\zeta}}}\,{dp\over d\zeta}\,\Big)\,\eta_{\mu\nu}
\ , \ \mu,\nu=0,1,2 \\ 
 T_{\zeta\zeta}^{branes} & = & -3\,Q_f\,\big(g_{\zeta\zeta}\big)^{{3\over 2}}\,e^{3\lambda+8\gamma+{3\phi\over 2}}\,\,p\ .
\eea
One can readily verify that our background satisfies (\ref{5d_scalar_eoms_compact}) and (\ref{Einstein_eqs_5d}) for an arbitrary profile function $p(\zeta)$.

\section{Degrees of anisotropy}\label{app:Deg_ani}

The effective Lifshitz exponent (\ref{z_eff_p_phi_f})  can  be written in terms of the master function $W$ as:
\be\label{z_eff_W}
{1\over z_{eff}}\,=\,1\,-\,{Q_f\over W\,+\,{1\over 6}\,\zeta\,{d W\over d\zeta} }\,{p\over \zeta\,\sqrt{W}}\,=\,1\,+\,\zeta\,{d\over d\zeta}\,\log\Big[W\,+\,{1\over 6}\,\zeta\,{d W\over d\zeta} \Big]\ .
\ee
When the master function is given by (\ref{W_general_sol}), the effective exponent depends on two integers $n$ and $m$ and can be written as
\be
 z_{eff}= {1+{Q_f\over 4}\,(\kappa\zeta)^{-4}\,  F\Big({4\over m}, {3+n\over m}; {4+m\over m};- (\kappa \zeta)^{-m}\Big)\over   1-Q_f\,(\kappa\zeta)^{-4}\,\Big[
  \big(1+(\kappa \zeta)^m\big)^{-{3+n\over m}}\,+{1\over 4}  F\Big({4\over m}, {3+n\over m}; {4+m\over m};- (\kappa \zeta)^{-m}\Big) \Big]
  }\ . 
\ee
From this expression we can readily obtain the behavior (\ref{zeff_UV}) of $z_{eff}$ in the UV region $\zeta\to\infty$.
In order to obtain the behavior of $z_{eff}$ as $\zeta\to 0$ it is convenient to rewrite  $z_{eff}$ as
\be
 z_{eff}={1-\,{Q_f\,(\kappa\zeta)^{n-1}\over (n-1)\,w_{n,m}} F\Big({n-1\over m}, {3+n\over m}; {n+m-1\over m};- (\kappa \zeta)^m\Big)\over  1-{Q_f\,(\kappa\zeta)^{n-1}\over \,w_{n,m}}\,\big(1+(\kappa \zeta)^m\big)^{-{3+n\over m}}\, \big[1+{1\over n-1}\,\big(1+(\kappa \zeta)^m\big)  F\big(1, {m-4\over m}; {n+m-1\over m};- (\kappa \zeta)^m\big)\Big] }\ .
\ee
The IR behavior for $z_{eff}$ for both boomerang and anisotropic Lifshitz flows can be readily obtained from this last equation, resulting in (\ref{eq:zeff_IR}).

\subsection{The internal squashing function}

The D5-brane sources cause both the anisotropy of the model and the deformation of the internal manifold. The latter is most conveniently characterized by the so-called internal squashing function $q=q(\zeta)$, defined as
\be
 q(\zeta)\equiv {e^{f(\zeta)}\over \zeta}\ .
\ee
This measures the deviation of the internal metric from that of the round ${\mathbb S}^5$. It takes a simple form in terms of the dilaton and the master function $W$, and can also be written entirely using the master function
\be\label{q_W}
 q = \sqrt{e^{\phi}\,W} = \frac{1}{\sqrt{1+{1\over 6}\,\zeta\,{d\log W\over d\zeta}}} \ .
\ee
From the latter it is rather easy to obtain the asymptotic forms of $q$. In the UV,
\be
 q = 1+{Q_f\over 4\,(\kappa\zeta)^4}+\ldots\ , \ \zeta\to\infty \ .
\ee
For Lifshitz solutions $q$ attains a constant value in the IR that depends on $n$, while for the boomerang solutions the ${\mathbb S}^5$ rounds out again, $\zeta\to 0$,
\be
 q = \left\{ \begin{array}{ll}
    1+{Q_f\over 2(n+5)w_{n,m}}(\kappa\zeta)^{n-1} + \ldots & , \ \ n>1 \\
    \sqrt{{6\over n+5}} + \ldots & , \ \ n<1 \ .\end{array}\right. 
\ee

\begin{figure}[ht]
\center
 \includegraphics[width=0.48\textwidth]{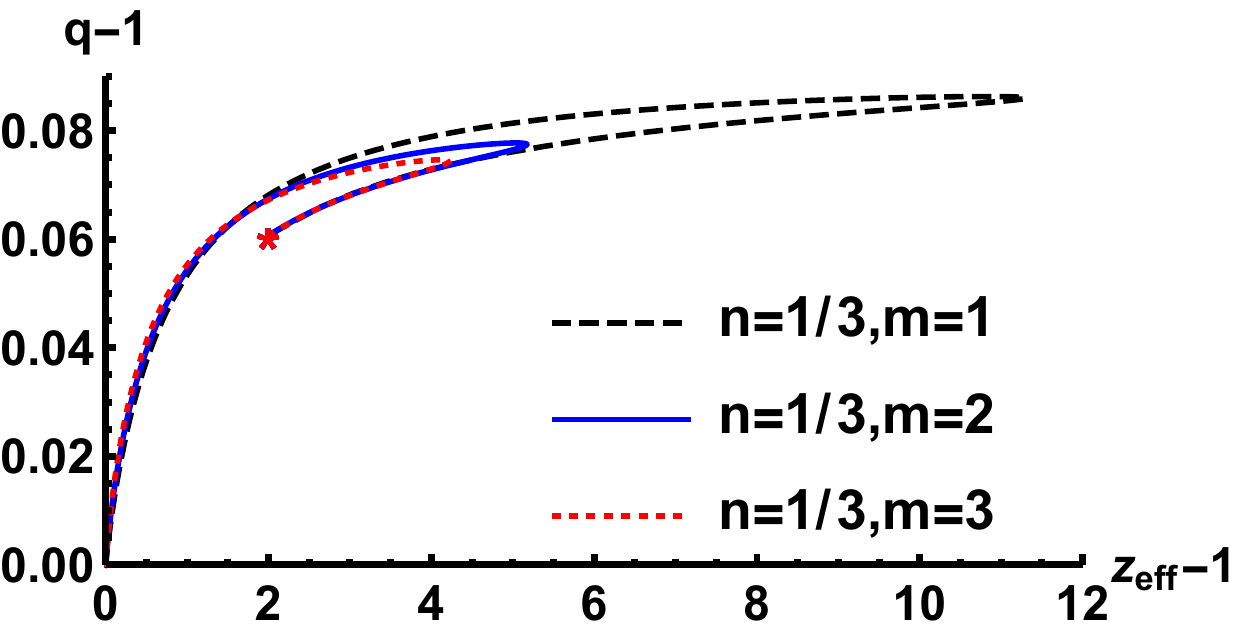}
 \includegraphics[width=0.48\textwidth]{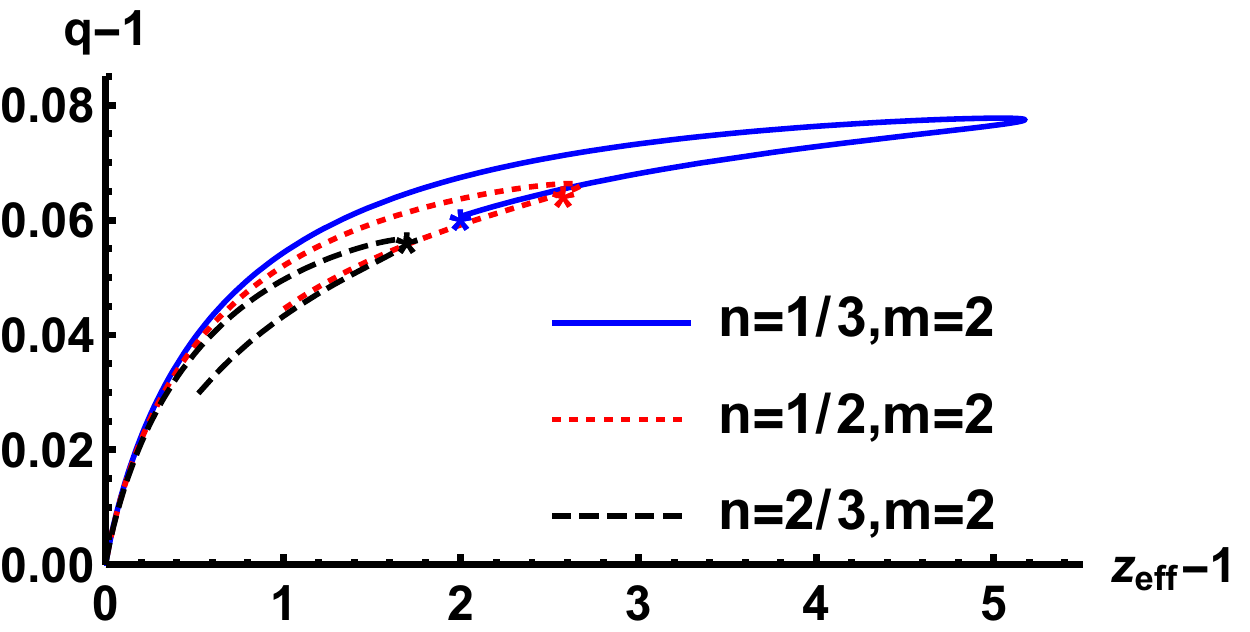}
  \caption{We present the internal squashing function versus the effective exponent for Lifshitz flows. Left: The curves correspond to fixed $n=1/3$ and $m=1$ (dashed black), $m=2$ (blue), and $m=3$ (dotted red). Right: The curves correspond to fixed $m=2$ and varying $n=1/3$ (blue), $n=1/2$ (dotted red), and $n=2/3$ (dashed black). In both panels we have fixed $Q_f=1$. The RG flows clockwise.}
\label{q_z_Lifshitz}
\end{figure}

\begin{figure}[ht]
\center
 \includegraphics[width=0.48\textwidth]{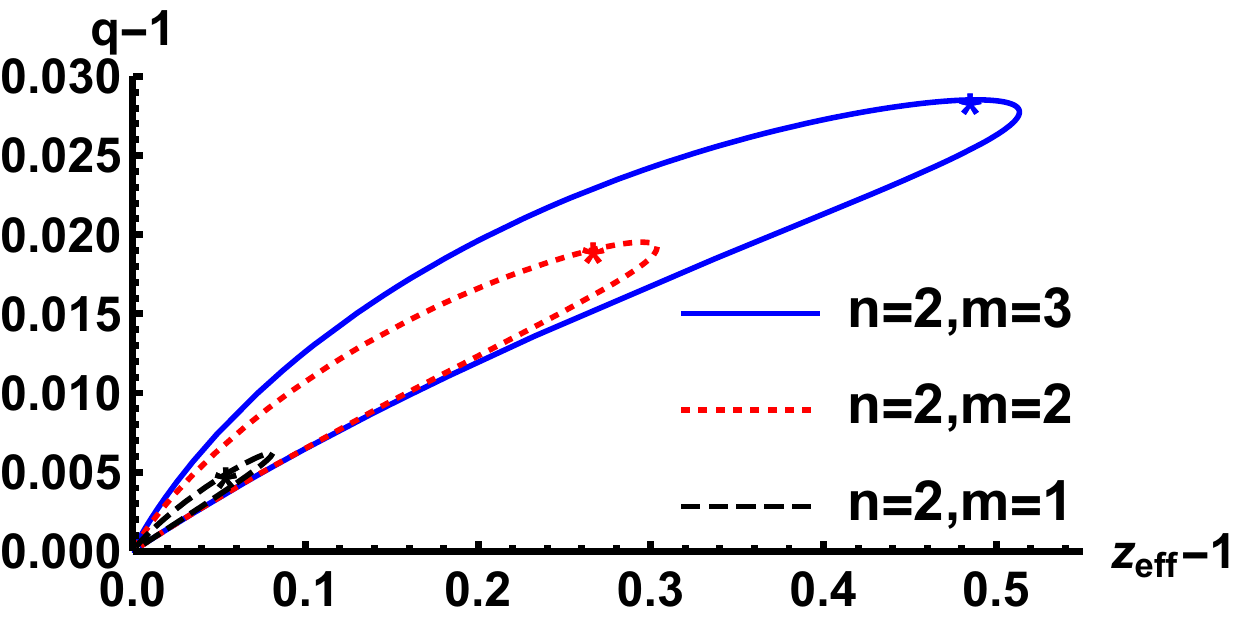}
 \includegraphics[width=0.48\textwidth]{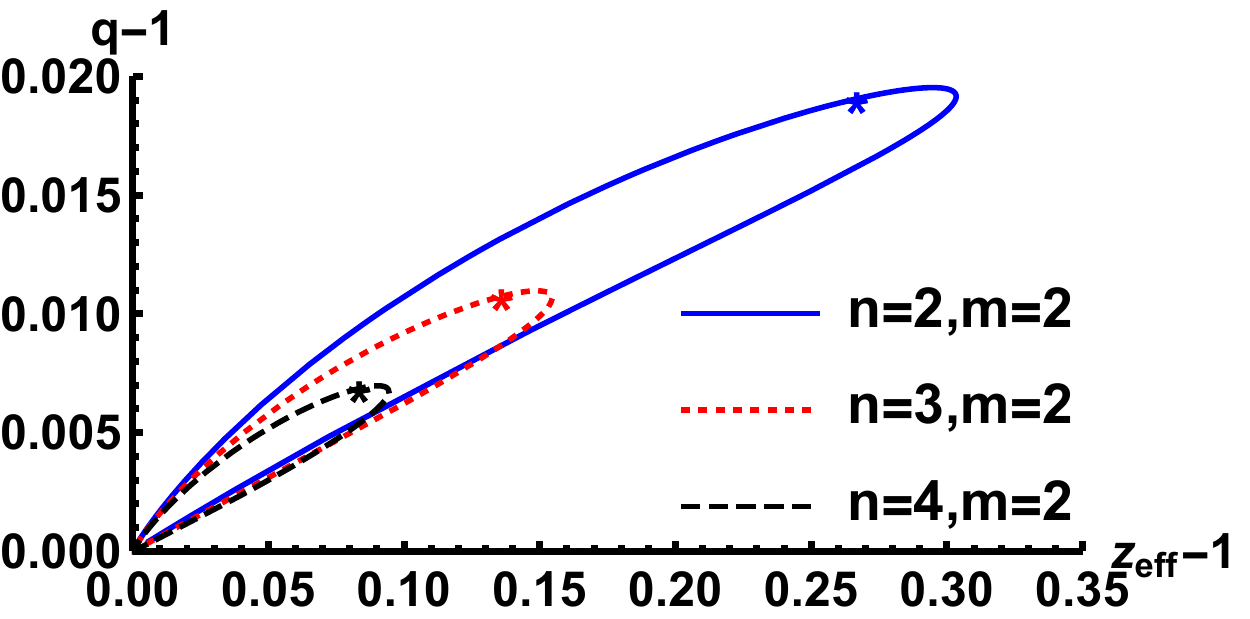}
  \caption{We present the internal squashing function versus the effective exponent for boomerang flows. Left: The curves correspond to fixed $n=2$ and $m=1$ (blue), $m=2$ (dotted red), and $m=3$ (dashed black). Right: The curves correspond to fixed $m=2$ and varying $n=2$ (blue), $n=3$ (dotted red), and $n=4$ (dashed black). In both panels we have fixed $Q_f=1$. The RG flows clockwise.}
\label{q_z_Boomerang}
\end{figure}

By numerical investigation one finds that $q(\zeta)$ resembles $z_{eff}$ very closely. The deviations from the round ${\mathbb S}^5$ are maximal at roughly the same values of $\zeta$ where $z_{eff}$ is also maximal. A natural question then arises if there is a simple relation between $ z_{eff}$ and $q$. One can find this relation by appropriately subtracting (\ref{z_eff_W}) from  (\ref{q_W}):
\be
 {1\over z_{eff}}-{1\over q^2}=\zeta\,{d\over d\,\zeta}\,\log {W\,+\,{1\over 6}\,\zeta\,{d W\over d\zeta} \over W^{{1\over 6}}}\ .
\ee
In order to get further insight on the relation between these two functions we have plotted $q$ versus $z_{eff}$ for Lifshitz (Fig.~\ref{q_z_Lifshitz}) and boomerang (Fig.~\ref{q_z_Boomerang})  flows.  The $q(z_{eff})$ curves are double-valued and have the shape of a lasso. The upper (lower) portion of the $q(z_{eff})$ corresponds to the UV (IR) region, whereas the turning point corresponds roughly to the value of $\zeta$ where the anisotropy is maximal. In other words, the flows from the UV to the IR correspond to clockwise paths. In the boomerang solutions the  $q(z_{eff})$ curve is closed. This is not the case for Lifshitz geometries since $z_{eff}\not\to 1$ as $\zeta\to 0$.

\section{Calculation of the Entanglement Entropy}\label{app:EE}

Let us fill in some background details in the computation of the holographic entanglement entropy. The holographic entanglement entropy of a strip consisting of two flat surfaces separated along a general spatial direction $x$ reads
\be
 S_E=\frac{1}{4 G_{10}}\int d^8\sigma \sqrt{g_8} \ ,
\ee
where the induced metric is 
\be
 ds_8^2 = g_{ij}dy^i dy^j+\left(g_{\zeta\zeta}+g_{xx}(x')^2\right)d\zeta^2+g_S ds_{\mathbb{CP}^2} +g_{\tau\tau}(d\tau+A)^2\ .
\ee
The $y^i$, $i=1,2$ are the coordinates parallel to the boundaries of the strip in the field theory directions. Explicitly,
\be
 S_{EE} = \frac{\pi^3}{4 G_{10}}\int d^2 y d\zeta g_S^2\left(g_{\tau\tau}g_{11}g_{22}g_{\zeta\zeta}\right)^{1/2}\sqrt{1+\frac{g_{xx}}{g_{\zeta\zeta}}x'^2} \ .
\ee
We consider a now a generic minimal surface anchored at the boundary on straight lines separated along the $x$ direction. The surface will have a profile $x(\zeta)$, and the area is given by
\be
 S = c\int d\zeta A \sqrt{1+B x'^2} \ .
\ee
In this expression $A$, $B$ are functions of $\zeta$ and $c$ a constant. We assume that the boundary is at $\zeta = \infty$. As usual with strip configurations, there is a first integral due to  conjugate momentum being independent of $\zeta$,
\be
 \frac{\delta S}{\delta x'} = -c P \ .
\ee
This gives a configuration of width $\ell$ that extremizes the area
\be\label{eq:Lgen}
 x'=-\frac{P}{A B} \frac{1}{\sqrt{1- \frac{P^2}{A^2 B}}} \ , \ \ell=\int d\zeta \, x' \ .
\ee
The action evaluated on the extremal configuration is 
\be\label{eq:Sgen}
 S=c\int d\zeta \frac{A}{\sqrt{1- \frac{P^2}{A^2 B}}} \ .
\ee

Let us now focus on our background and infer the data going into the above formulas:
\be\label{eq:ABfuncs}
 c=\frac{\pi^3}{4 G_{10}} \ , \ A=g_S^2\left(g_{\tau\tau}g_{11}g_{22}g_{\zeta\zeta}\right)^{1/2} \ , \ B=\frac{g_{xx}}{g_{\zeta\zeta}} \ .
\ee
In all the cases we have that
\be
g_S=h^{1/2}\zeta^2 \ , \ g_{\tau\tau}=h^{1/2}e^{2f} \ ,\ g_{\zeta\zeta}=h^{1/2}\zeta^2 e^{-2f} \ .
\ee
For the other components we have the following options
\begin{itemize}
\item $x$ parallel to the anisotropic direction
\be
g_{11}=g_{22}=h^{-1/2}\ ,\ g_{xx}= h^{-1/2}e^{-2\phi}\ .
\ee
\item $x$ transverse to the anisotropic direction
\be
g_{xx}=g_{11}=h^{-1/2}\ , \ g_{22}= h^{-1/2}e^{-2\phi} \ .
\ee
\end{itemize}
Then, the coefficients are
\begin{itemize}
\item $x$ parallel to the anisotropic direction
\be
A=\zeta^5 h\ , \ B^{-1}=\zeta^2 h e^{2\phi-2f}\ , \ A^{-2}B^{-1}=\zeta^{-8} h^{-1}e^{2\phi-2f} \ .
\ee
\item $x$ transverse to the anisotropic direction
\be
A=\zeta^5 h e^{-\phi}\ , \ B^{-1}=\zeta^2 h e^{-2f}\ , \ A^{-2}B^{-1}=\zeta^{-8} h^{-1}e^{2\phi-2f} \ .
\ee
\end{itemize}
We define $\zeta_0$ as the position at the bottom of the surface, which is the solution to the equation $\zeta_0^8=P^2 h^{-1}e^{2\phi-2f}\Big|_{\zeta=\zeta_0}$. We introduce a cutoff in the radial direction $\zeta_\Lambda$. 

From the formulas above, the entanglement entropy \eqref{eq:see} and the width of the strip \eqref{eq:ell} directly follow.  Close to the boundary, where $h\sim R_{UV}^4/\zeta^4$, $e^{2f}\sim \zeta^2$,
\be
 S_{EE}\sim \frac{\pi^3 V_2}{2 G_{10}}R_{UV}^4\int d\zeta \zeta\left( 1+O(\zeta^{-4})\right) \ .
\ee
There is a quadratic UV divergence, we will subtract it to get the finite part of the entropy, which we denote as $\hat{S}_{EE}$.

\subsection{UV asymptotics}

We start with \eqref{eq:see} and the asymptotic UV expansions
\be
h\simeq \frac{R_{UV}^4}{\zeta^4}\ , \ e^{-\phi}\simeq 1 \ , \ e^{2 f}\simeq \zeta^2 \ .
\ee
We will use the condition that relates the constant $P$ with the tip of the entangling surface $\zeta_0$, $\zeta_0^8\simeq \frac{P^2}{R_{UV}^4}\zeta_0^2$.
Then we find $P\simeq R_{UV}^2 \zeta_0^3$. We will do an expansion in $\zeta,\zeta_0\to \infty$ with $\zeta_0/\zeta$ fixed. The term inside the square root goes as 
\be
 P^2\frac{e^{2\phi-2f}}{\zeta^8 h}\sim \frac{\zeta_0^6}{\zeta^6} \ .
\ee
At leading order the expansion of the integrands in the entropy are
\be
 \sim \zeta  \left(\frac{1}{\sqrt{1-\frac{\zeta _0^6}{\zeta ^6}}}-1\right) R_{UV}^4 \ .
\ee

In order to compute the integrals we will change variables to $\zeta=\zeta_0 u^{-1/6}$ and integrate $u\in [0,1)$. Denoting $s_0=\frac{\pi^3 V_2}{2 G_{10}}$, as $\zeta_0\to \infty$, there is a leading contribution proportional to a coefficient
\be
 c_0=\frac{\sqrt{\pi}\Gamma\left(\frac{2}{3}\right)}{2\Gamma\left(\frac{1}{6}\right)} \ .
\ee
We can approximate the regulated entanglement entropy by
\be
 \hat{S}_{EE}^\parallel \simeq \hat{S}_{EE}^\perp \simeq -s_0 c_0 R_{UV}^4 \zeta_0^2 \ .
\ee
The separation between the two walls have integrands that go as
\be
 \sim \frac{1}{\zeta ^5 \sqrt{1-\frac{\zeta _0^6}{\zeta ^6}}} \ .
\ee
Computing the integrals, substituting the value of $P$, and expanding one finds that the first term is proportional to the coefficient $4 c_0$, allowing us to solve for $\zeta_0$:
\be
 \ell\simeq 4 c_0 \frac{R_{UV}^2}{\zeta_0} \ \rightarrow \ \zeta_0\simeq 4 c_0\frac{R_{UV}^2}{\ell} \ .
\ee
Plugging this in the expressions for the entanglement entropy and expanding we find
\be
 \hat{S}_{EE}\simeq-16 c_0^3\frac{s_0R_{UV}^8}{\ell^2} \ .
\ee

It is straightforward, albeit a bit longer, derivation to get the subsubleading behaviors at the UV. We are content with representing the final result of the UV expansion to the next order:
\be\label{eq:UVsubsub}
 \hat{S}_{EE}^{\parallel,\perp}\simeq-16 c_0^3\frac{s_0R_{UV}^8}{\ell^2}\left(1-\gamma_{\parallel,\perp} \frac{Q_f}{8(\kappa \ruv)^4}\left(\frac{\ell}{\ruv}\right)^4 \right) \ ,
\ee
where
\be
 \gamma_\parallel = \frac{2}{5}\gamma_\perp = \frac{\Gamma(1/6)^7}{120\times 2^{2/3}\pi^{7/2}\Gamma(2/3)^4} \ .
\ee
We have checked this asymptotic result against the numerical calculation, see Fig.~\ref{fig:boomerangcfunction}.

\subsection{IR asymptotics}

We will separate the finite part of the entanglement entropy in an IR contribution and a UV contribution, separated by some scale $\zeta_M$. The IR contribution is obtained by integration up to $\zeta_M$. The approximate expressions depend on the IR behavior. 

For the boomerang flows, the expansions are essentially the same as in the UV, except for the anisotropic coordinate, which has an additional constant scale factor. In the calculation, the functions $A$, $B$ in \eqref{eq:ABfuncs} change relative to the UV case by a factor
\begin{itemize}
\item $x$ parallel to the anisotropic direction
\be
A\to A \ , \ B\to w_{n,m}^2B \ .
\ee
\item $x$ transverse to the anisotropic direction
\be
A\to w_{n,m}A \ , \ B\to B \ .
\ee
\end{itemize}
Recall, that $w_{n,m}$ is given in \eqref{eq:wnm}. The dependence on $w_{n,m}$ can be removed from inside the square root in \eqref{eq:Sgen} by rescaling $P$
\be
 P\to w_{n,m}P \ .
\ee
The combination of all these rescalings introduce the following factors in the EE and the width
\bea
 \hat{S}_{EE}^\parallel & \to & \hat{S}_{EE,UV}^\parallel(\zeta_0),\ \ \ell_\parallel\to w_{n,m}^{-1}\ell_{UV,\parallel}(\zeta_0) \\
 \hat{S}_{EE}^\perp & \to & w_{n,m}\hat{S}_{EE,UV}^\perp(\zeta_0),\ \ \ell_\perp\to \ell_{UV,\perp}(\zeta_0) \ .
\eea
From these, it is easy to derive \eqref{eq:seeboom}. 
The next order correction follows from the expansion \eqref{eq:irexpboom}. For $5>n>1$ the scaling in all directions is
\be
\ell \sim \frac{1}{\zeta_0}\left(a+Q_f b\left(\frac{\zeta_0}{\zeta_m}\right)^{n-1}\right).
\ee
Since the scaling does not depend on the direction we have dropped the label, but one should keep in mind that the coefficients are different in each direction. We have introduced $\zeta_m$ to fix the units, which should be a characteristic scale of the background geometry. The value of $\zeta_m$ or $b$  cannot be determined just from the IR geometry, but the full profile is needed. For $n>5$ the power of the NLO correction inside the bracket remains at a value of $4$, independently of the value of $n$. The EE also has similar scalings in all the directions, for $5>n>1$,
\be
\hat{S}_{EE} \sim \zeta_0^2\left( c+Q_f d\left(\frac{\zeta_0}{\zeta_m}\right)^{n-1} \right) \ ,
\ee
where again the coefficients $c$ and $d$ depend on the direction, even if the scaling does not. Solving for $\zeta_0$ in terms of $\ell$ and plugging the result in the EE one finds
\be
\hat{S}_{EE}\sim \frac{4 a}{\ell^2}\left(a c+Q_f\left( 2^{n-1}a d+ (n-1)  b c\right) \left(\frac{\ell_m}{\ell_0}\right)^{n-1} \right) \ , \ \ell_m=a/\zeta_m \ .
\ee

Finally let us discuss the case $n<1$. For geometries with anisotropic Lifshitz scaling,
\bea
\hat{S}_{EE}^\parallel & = & \frac{\pi^3 V_2}{2 G_{10}}\int_{\zeta_0}^{\zeta_M} d\zeta\frac{R^4}{\lambda_n^4}\frac{\zeta }{\sqrt{1-\frac{\lambda_n^6 P^2}{R^4 \zeta^6(\mu \zeta)^{2(n-1)}}}}+\ldots \\
\hat{S}_{EE}^\perp & = & \frac{\pi^3 V_2}{2 G_{10}}\int_{\zeta_0}^{\zeta_M} d\zeta \frac{R^4}{\lambda_n^4} \frac{\zeta (\mu \zeta)^{n-1}}{\sqrt{1-\frac{\lambda_n^6 P^2}{R^4 \zeta^6(\mu \zeta)^{2(n-1)}}}}+\ldots \ .
\eea
The constant $\lambda_n$ was defined in \eqref{eq:lncn}. In this case $R^4\zeta_0^6(\mu \zeta_0)^{2(n-1)}=\lambda_n^6 P^2$. The expressions for the length are in each case
\bea
 \ell_\parallel & = &  2\lambda_n^2P\int_{\zeta_0}^{\zeta_M} \frac{d\zeta}{\zeta^5} \frac{(\mu \zeta)^{2(1-n)}}{\sqrt{1-\frac{\lambda_n^6 P^2}{R^4 \zeta^6(\mu \zeta)^{2(n-1)}}}} +\ldots\\
 \ell_\perp & = & 2\lambda_n^2 P\int_{\zeta_0}^{\zeta_M} \frac{d\zeta}{\zeta^5}  \frac{(\mu \zeta)^{1-n}}{\sqrt{1-\frac{\lambda_n^6 P^2}{R^4 \zeta^6(\mu \zeta)^{2(n-1)}}}} +\ldots \ .
\eea

The integrals can be calculated explicitly in terms of Gamma and Beta functions. Expanding for small values of $\zeta_0$, one finds the leading order behavior for the entanglement entropy to be
\bea
 \hat{S}_{EE}^\parallel & \sim & -\frac{\pi^3 V_2}{2 G_{10}} R^4A_\parallel\zeta_0^2 \\
 \hat{S}_{EE}^\perp & \sim & -\frac{\pi^3 V_2}{2 G_{10}} R^4 A_\perp\zeta_0^2 (\mu \zeta_0)^{n-1} \ .
\eea
The coefficients $A$ are given in \eqref{eq:AB}. The separation between the two walls is
\bea
 \ell_\parallel & \sim & 4\lambda_n^2P\mu^{2(1-n)} \frac{c_{n-1}^\parallel}{n}\zeta_0^{-2(n+1)} =R^2 \mu^{1-n} B_\parallel \zeta_0^{-n} \\
 \ell_\perp & \sim & 2(n+1)\lambda_n^2 P\mu^{1-n} c_{n-1}^\perp \zeta_0^{-n-3}=R^2 B_\perp \zeta_0^{-1} \ ,
\eea
where the different coefficients can be found in \eqref{eq:lncn} and \eqref{eq:AB}. Therefore, solving for $\zeta_0$ in terms of $\ell$ and plugging the result in to the entanglement entropy, the asymptotic behaviors of the entanglement entropy with the separation between the walls as given by \eqref{eq:seescale} follows.

\bibliographystyle{JHEP}
\bibliography{refs}

\end{document}